\documentclass{article}
\usepackage{cite}
\usepackage{amsmath,amssymb,amsfonts}
\usepackage{algorithmic}
\usepackage{textcomp}

\usepackage[pdftex]{graphicx}	

\usepackage{here}
\usepackage{graphicx}
\usepackage{latexsym}
\usepackage{comment}
\usepackage{multirow}
\usepackage{bm}
\usepackage{cite}
\usepackage[hang,small,bf]{caption}
\usepackage[subrefformat=parens]{subcaption}
\def\lla{\left\langle}
\def\rra{\right\rangle}
\def\lp{\left(}
\def\rp{\right)}

\def\graphsize{0.400}
\def\graphsizeB{0.600}

\usepackage{bm}

\title{Exact equality of the MSEs for two types of nonlinear adaptive systems: 
Saturation and dead-zone types}
\author{
Seiji~Miyoshi\thanks{Department of Electrical, Electronic and Information Engineering, 
Faculty of Engineering Science, 
Kansai University, 
3-3-35 Yamate-cho, Suita-shi, Osaka 564-8680, Japan
(e-mail: miyoshi@kansai-u.ac.jp),
This work was partially supported by JSPS KAKENHI 
Grant Number JP20K04494.}
and
Yua~Yamaguchi\thanks{Nidec Advance Technology Corporation,
Nidec PARK Bldg. C, 1--1 Higashinokuchi, Morimoto-cho, Muko-shi, 
Kyoto 617-0003, Japan}}

\begin{document}

\maketitle

\section*{Abstract}
Adaptive signal processing systems, commonly utilized in applications 
such as active noise control and acoustic echo cancellation, often 
encompass nonlinearities 
due to hardware components such as loudspeakers, microphones, and amplifiers.
Examining the impact of these nonlinearities on 
the overall performance of adaptive systems is critically important.
In this study, we employ a statistical-mechanical method to investigate the behaviors 
of adaptive systems, each containing an unknown system 
with a nonlinearity in its output.
We specifically address 
two types of nonlinearity: saturation and dead-zone types. 
We analyze both the dynamic and steady-state behaviors of 
these systems under the effect of such nonlinearities. 
Our findings indicate 
that when the saturation value is equal to the dead-zone width, 
the mean square errors (MSEs) in steady states are identical for 
both nonlinearity types. 
Furthermore, we derive a self-consistent equation to obtain 
the saturation value and dead-zone width that maximize the steady-state MSE.
We theoretically clarify
that these values depend on neither the step size nor 
the variance of background noise.

\bigskip

\noindent
Keywords: adaptive filter, 
adaptive signal processing, 
system identification, 
LMS algorithm, 
saturation-type nonlinearity,
dead-zone-type nonlinearity, 
statistical-mechanical analysis

\section{Introduction} \label{sec:introduction}
Adaptive signal processing is utilized extensively 
in fields such as communication and acoustic systems\cite{Haykin2002, Sayed2003}. 
It plays a key role in various applications including 
active noise control (ANC)\cite{Nelson1992,Kuo1996,Kuo1999,Kajikawa2012}, 
active vibration control (AVC)\cite{Fuller1995}, 
acoustic echo cancellation\cite{Sondhi2006}, 
and system identification\cite{Ljung1999}. 
These systems often include power amplifiers and transducers, 
such as loudspeakers and microphones, both of which exhibit 
significant nonlinearities\cite{Haykin2002, Sayed2003}. 
Thoroughly examining the impact of these elements on 
the overall performance of adaptive systems is critically important.
Therefore, numerous studies
\cite{Widrow1971,Claasen1981,Duttweiler1982,Aref2015,Smaoui2011,Bekrani2014,Deivasigamani1982,Bekrani2012,Jun1995,Takahashi1992,Eweda1990,Bershad1990,Costa2001,Costa1999,Costa2002,Snyder1995,Costa2017,Costa2008,Tobias2006,Bershad2009,Hamidi2004,Stenger2000,Miyoshi2013,Motonaka2021,Kugiyama2024,Miyoshi2011EL,Miyoshi2018,Mathews1991,Miyoshi2022}
have been conducted on 
adaptive signal processing systems that integrate these nonlinear elements
as discussed in detail in Sect. \ref{sec:related}.

Thus, the analysis of adaptive signal processing systems involving 
nonlinearities is extremely important, and as the unknown system and 
the adaptive filter are the two main components in 
an adaptive signal processing system, 
the analysis of their nonlinearities is equally important.
However, as will be discussed in Sect. \ref{sec:related}, 
there has been much research on the former, i.e., 
the nonlinearity on the adaptive filter side, which corresponds, for example, 
to the model of an ANC, but very little on the latter, i.e., 
the nonlinearity on the unknown system side, which corresponds, 
for example, to the model of an acoustic echo canceller.
%
Therefore, it is extremely significant to clarify the behavior of adaptive signal processing systems with nonlinearity on the unknown system side.
This is the motivation for this study, 
however,
note 
that the underlying motivation is 
the desire to deeply and comprehensively 
understand and appreciate the complex behaviors 
of adaptive signal processing systems themselves, including nonlinear elements.
In this study, a statistical-mechanical method is 
used for the theoretical analysis.
Two types of nonlinearity in 
un unknown system are investigated: 
saturation-type nonlinearity and dead-zone-type nonlinearity.
The main contributions of this paper are as follows:

\begin{itemize}
\item 
We explore 
the behavior of an adaptive signal processing system 
where the output of the unknown system exhibits saturation-type nonlinearity, 
and separately, where it exhibits dead-zone-type nonlinearity 
by applying a statistical-mechanical method.
The work conducted by Miyoshi\cite{Miyoshi2022} focused 
on an active noise control model where nonlinearity 
was present in the adaptive filter.
In contrast, our current study 
shifts its focus to an acoustic 
echo canceller model, which is characterized by the presence of 
nonlinearity in the unknown system, not in the adaptive filter.
\item 
For the adaptive signal processing system 
in which the unknown system is modeled by the FIR filter
and the adaptive filter is composed of the FIR filter,
the dynamical and steady-state behaviors of 
the mean square error (MSE) and
the mean square deviation (MSD)
are discussed deterministically
under the assumptions of a long filter.
\item 
We compare the behaviors of saturation-type and dead-zone-type
nonlinearities.
As a result, we interestingly found 
that the steady-state MSEs for the saturation and dead-zone types 
are exactly the same.
In addition, 
we obtain the self-consistent equation for 
the saturation and dead-zone values
that maximize the steady-state MSEs.
Statistical-mechanical methods used by Miyoshi \cite{Miyoshi2022} and others are so powerful that they again yield entirely new and very interesting findings in this paper.
\end{itemize}
%
The remainder of this paper is structured as follows.
In Sect. \ref{sec:model}, we provide details of the model, which is 
the focus of our study, laying out the foundational 
concepts and parameters involved. 
In Sect. \ref{sec:analysis}, 
we describe a thorough statistical-mechanical analysis of the model, 
elaborating on the methodologies and techniques employed. 
In Sect. \ref{sec:results_and_discussion}, we present 
the core findings of our research, with which we validate 
the derived theory by comparing the theoretical 
predictions with numerical simulation data. Finally, 
in Sect. \ref{sec:conclusions}, we summarize the insights 
gained from our study and discuss the implications 
of our findings, concluding the paper with a summary 
of key outcomes and potential avenues for future research.

{\it Notation:}
Scalars are denoted by lowercase italic fonts
except for $Q$, $S$, $D$, $Z$, $N$, $M$, $A_S$, and $A_D$, which are also scalars
in accordance with the conventions used in the corresponding literature. 
Column vectors are denoted by bold lowercase italic fonts
and matrices by bold uppercase italic fonts. 
The superscripts ${}^\top$ and ${}^{-1}$ denote transpose and inverse, 
respectively, 
whereas $\langle \cdot \rangle$ stands for expectation.
Finally, if $\bm{z}$ is a column vector, 
then $\|\bm{z}\|_2^2=\bm{z}^{\top}\bm{z}$.

\section{Related Works} \label{sec:related}
As described in Sect. \ref{sec:introduction},
the analysis of adaptive signal processing systems involving 
nonlinearities is extremely important.
Therefore, numerous studies have been conducted on 
adaptive signal processing systems that integrate nonlinear elements.
Bershad\cite{Bershad1990} analyzed the effects of saturation-type nonlinearity in 
the least-mean-square (LMS) algorithm, specifically considering a small step size 
and a nonlinearity described by the formula $(1-e^{-ax})$. 
Costa {\it et al.}\cite{Costa2001} studied  
adaptive filters with error function (erf)-saturation-type nonlinearity,
assuming a small step size. 
Costa {\it et al.}\cite{Costa1999,Costa2002} analyzed
ANC in which the secondary path
has an erf-saturation-type nonlinearity.  
Furthermore, Snyder and Tanaka\cite{Snyder1995} 
suggested replacing the finite-duration impulse response (FIR) filter 
with a neural network to deal with the primary path nonlinearity in 
ANC and AVC systems. 
Costa\cite{Costa2017} conducted an analysis of 
a hearing aid feedback canceller that featured erf saturation-type nonlinearity.
Costa {\it et al.}\cite{Costa2008}
investigated a model in which the output of the adaptive filter
demonstrates dead-zone-type nonlinearity 
under the assumption of a small step size. 
This type of nonlinearity arises from class B amplifiers or 
nonlinear actuators.
Tobias and Seara\cite{Tobias2006} explored modifications of the LMS algorithm 
designed to improve its performance in environments affected by 
erf saturation-type nonlinearity. 
Bershad\cite{Bershad2009} analyzed 
how the LMS algorithm is updated when it encounters 
erf saturation-type nonlinearity. 
Furthermore, Bershad\cite{Bershad2009} expanded 
this analysis to include the tracking of a Markov channel, 
specifically within the realm of system identification.
As described thus far, numerous studies have been conducted 
on adaptive systems that include erf saturation-type nonlinearity.
On the other hand,
Hamidi {\it et al.}\cite{Hamidi2004} conducted an analysis, 
together with computer simulations and experimental studies, 
of an ANC model that incorporates an adaptive filter with 
clipping saturation-type nonlinearity. 
To increase the efficiency of the adaptive algorithm, 
they suggested altering the cost function to circumvent 
the use of this nonlinear region.
Stenger and Kellermann\cite{Stenger2000} 
proposed clipping-type preprocessing in 
adaptive echo cancellation to mitigate the impacts of nonlinear echo paths.

Our research group's recent exploration of using 
a statistical-mechanical method\cite{Nishimori2001} 
to analyze adaptive signal processing 
represents a significant advance in handling the complexities of such systems. 
Conventional statistical analysis often relies on various approximations 
and assumptions to compute expectations concerning the input signal, 
which is inherently a random variable. 
However, the application of 
statistical-mechanical analysis enables a shift from this norm.
Statistical-mechanical analysis 
enables the examination of universal properties of systems composed of 
numerous microscopic variables
by assuming the large-system limit. 
This approach simplifies discussions to 
a few macroscopic variables, enabling a macroscopic and deterministic analysis. 
The benefits of this method are underscored by the applicability of 
the law of large numbers and the central limit theorem, which simplify 
many of the calculations involved.
This method proves especially beneficial for analyzing signal processing tasks 
that incorporate adaptive filters with 
markedly long taps, which is typical 
in practical acoustic systems. 
Our research group's application\cite{Miyoshi2011EL, Miyoshi2018} of 
the statistical-mechanical method to 
the analysis of feed-forward ANC systems updated by 
the Filtered-X LMS (FXLMS) algorithm represents an advanced approach 
in the field of adaptive signal processing. 
Initially, our analyses focused 
on systems in which the primary path, secondary path, and adaptive filter 
were linear \cite{Miyoshi2011EL, Miyoshi2018}. 
By expanding this focus in recent work,
our group\cite{Miyoshi2013, Motonaka2021, Kugiyama2024} has 
ventured into more complex scenarios involving 
nonlinearities, specifically analyzing models with both the unknown 
system and adaptive filter incorporating 
the Volterra-type nonlinearity\cite{Mathews1991}.
Although Volterra filters inherently exhibit 
nonlinear characteristics, adapting the statistical-mechanical 
method originally developed for linear systems to these filters 
has proven feasible. 
This adaptation, however, was limited to 
Volterra filters of a specific order, indicating a constraint 
in the versatility of the methodology. Additionally, the method has yet 
to address simpler yet commonly encountered nonlinearities in 
practical adaptive systems, such as those of the saturation and dead-zone types. 
These limitations highlight potential areas for further research and 
development within our group's work, aiming to expand 
the applicability of the statistical-mechanical method to 
a broader range of nonlinear scenarios encountered in adaptive signal processing.

Nonlinear components within 
adaptive signal processing systems have been extensively explored, 
particularly regarding the saturation characteristics inherent 
in power amplifiers and transducers such as loudspeakers 
and microphones. 
The erf saturation-type 
nonlinearity has been thoroughly studied and well understood 
owing to its mathematical tractability. However, another critical 
form of nonlinearity, the clipping saturation type, 
which exhibits piecewise linear characteristics, has not been as extensively analyzed.
Clipping saturation-type nonlinearity is crucial for 
accurately representing the saturation phenomena 
in adaptive systems. 
This type of nonlinearity is 
particularly challenging to model and analyze owing to 
its nondifferentiability. 
Costa {\it et al.} highlighted this 
in their study\cite{Costa2008}, in which they noted 
that to facilitate the development of analytical models, 
it is convenient to approximate the piecewise nonlinearity by
a continuous and more mathematically tractable function.
They\cite{Costa2008} even validated 
the theoretical results of their erf-type nonlinearity 
with computer simulations 
that employed piecewise linearity, 
underscoring the practical relevance of such studies.
Despite their importance, there remained a significant gap 
in the literature concerning the analytical study of 
clipping saturation-type nonlinearity. 
This gap indicated a potential area for future research, 
where more detailed analytical methods are needed to be developed to 
better understand and predict the behavior of 
adaptive systems encountering this type of nonlinearity.

%
On the basis of the aforementioned background,
Miyoshi\cite{Miyoshi2022}  investigated the behaviors of 
an adaptive system where the output of the adaptive filter 
exhibits clipping saturation-type nonlinearity using 
a statistical-mechanical approach.
His analysis clarified the existence of 
a critical saturation value at which 
the system toggles between mean-square stability and instability.
He also derived the exact value of this critical saturation.

\section{Model} \label{sec:model}
\begin{figure}[tbp]
\centering
\includegraphics[width=0.550\linewidth,keepaspectratio]{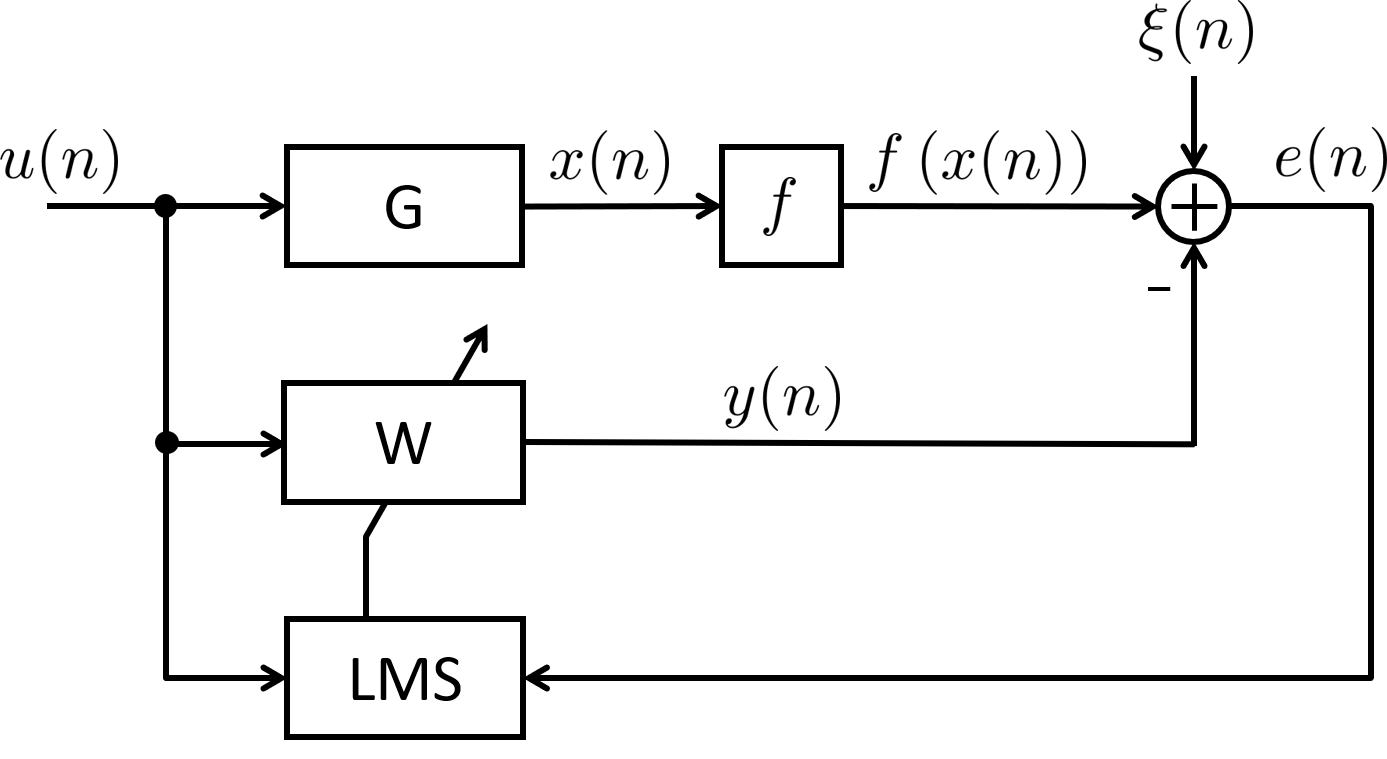}
\caption{Block diagram of the adaptive system.} \label{fig:block}
\end{figure}

Fig. \ref{fig:block} illustrates a block diagram of 
the adaptive system studied.
The impulse response of the unknown system G is an 
$M$-dimensional arbitrary vector
\begin{align}
\bm{g}_0&=[g_1,g_2,\ldots,g_M]^\top,
\end{align}
and remains time-invariant.
%
The adaptive filter W is an $N$-tap FIR filter,
characterized by its coefficient vector 
\begin{align}
\bm{w}(n)&=[w_1(n),w_2(n),\ldots,w_N(n)]^\top,
\end{align}
where $n$ denotes the time step.
%
Although the dimension $M$ of $\bm{g}_0$ is generally unknown
in advance, 
we assume that 
the tap length $N$ of the adaptive filter W is set to satisfy
\begin{align}
N\geq M,
\end{align}
because it is straightforward to design 
an adaptive filter W of tap length $N$ with a margin.
Moreover, let $\bm{g}$ be a vector expanded to $N$ dimensions 
by appending $N-M$ zeros to $\bm{g}_0$. That is,
\begin{align}
\bm{g}&=[g_1,g_2,\ldots,g_M,g_{M+1},\ldots,g_N]^\top,\\
g_i&=0, \ \ i=M+1,\ldots,N.
\end{align}
Note that although previous 
studies\cite{Costa2001,Costa2008,Tobias2006,Bershad2009}
typically assume that the dimensions of $\bm{g}_0$ and $\bm{w}$ 
are equal, our model diverges from this assumption. 
It permits arbitrary $\bm{g}_0$ dimensions and refrains from 
imposing strict constraints on its dimension $M$ or its elements ${g_i}, i=1,\ldots,M$.
We define the parameter $\sigma_g^2$ as
\begin{align}
\sigma_g^2&\triangleq \frac{1}{N}\|\bm{g}_0\|_2^2=\frac{1}{N}\|\bm{g}\|_2^2=
\frac{1}{N}\sum_{i=1}^N g_i^2.
\label{eqn:sigmag2}
\end{align}
As will become clear later, 
our theory relies solely on the unknown system G 
through $\sigma_g^2$. To validate this assertion, 
we will present empirical results in Sect. \ref{sec:results_and_discussion}, 
demonstrating the applicability of the theory to 
experimentally obtained $\bm{g}_0$.

The input signal $u(n)$
is assumed to be independently drawn from a distribution with
\begin{align}
\left\langle u(n) \right\rangle &=0, \hspace{10mm}
\left\langle u(n)^2 \right\rangle =\sigma^2.
\label{eqn:u_mean_variance}
\end{align}
That is, the input signal is white. 
Although the assumption of a white input signal may appear restrictive, 
it holds significant relevance in practical scenarios, 
notably in system identification. 
In addition, the analysis of white signals is crucial as a baseline.
Furthermore, this model provides valuable insights into algorithm behavior and 
serves as a benchmark for alternative cases\cite{Costa2001}.
Note that only the mean and variance of 
the distribution are specified in (\ref{eqn:u_mean_variance}). 
No specific distributions, such as the Gaussian distribution, 
are assumed.
The tap input vector $\bm{u}(n)$ at time step $n$ is
\begin{align}
\bm{u}(n)&=[u(n),u(n-1),\ldots,u(n-N+1)]^\top.
\end{align}

The expressions for $x(n)$ and $y(n)$ describe 
the outputs of the system G and the adaptive filter W, respectively, 
in terms of convolutions with their own coefficients 
and a sequence of input signals $u(n)$.
That is,
\begin{align}
x(n)&= \bm{g}^\top \bm{u}(n)   = \sum_{i=1}^N g_i    u(n-i+1), \label{eqn:d}\\
y(n)&= \bm{w}(n)^\top \bm{u}(n)= \sum_{i=1}^N w_i(n) u(n-i+1).  \label{eqn:y}
\end{align}

\begin{figure}[H]
\centering
\begin{minipage}[b]{0.48\linewidth}
\centering
\includegraphics[width=0.6\linewidth,keepaspectratio]{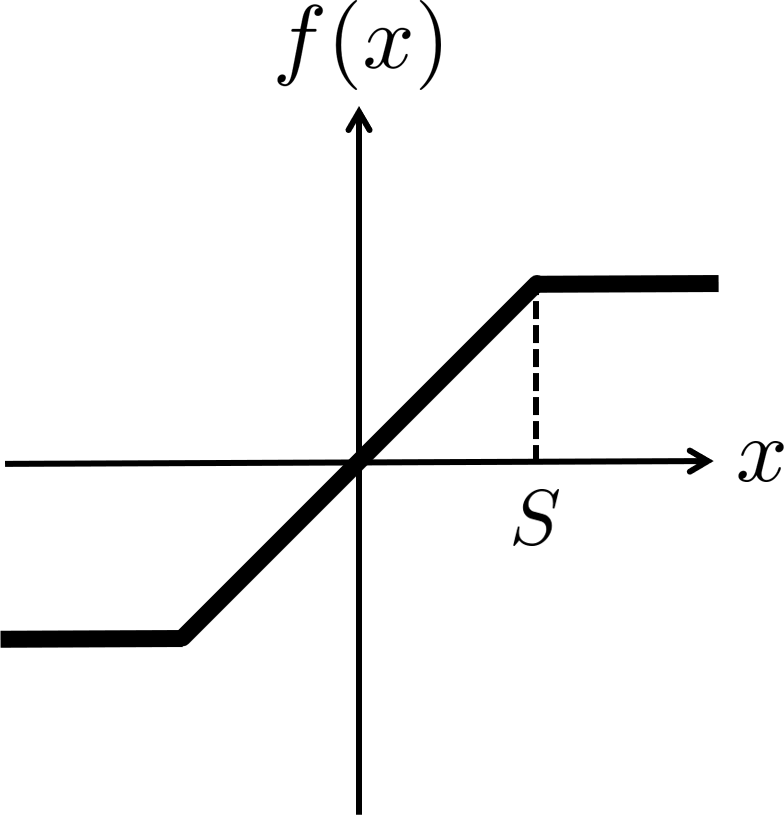}
\subcaption{Saturation type} \label{fig:saturation}
\end{minipage}
\begin{minipage}[b]{0.48\linewidth}
\centering
\includegraphics[width=0.6\linewidth,keepaspectratio]{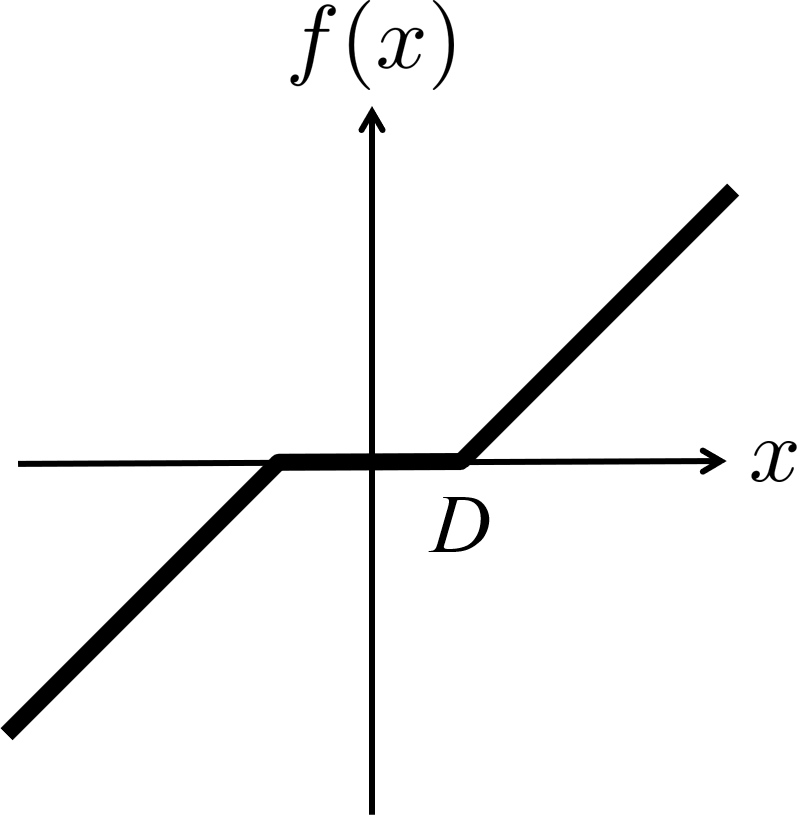}
\subcaption{Dead-zone type} \label{fig:deadzone}
\end{minipage}
\caption{Two types of nonlinearity}\label{fig:2nonlinearities}
\end{figure}

The nonlinearity of the unknown system G is modeled by
the function 
$f$ placed after G. 
In this paper, the function $f$ represents 
the saturation-type nonlinearity
\begin{align}
f(x)&=
	\begin{cases}
		S, 		& 	x>S\\
		-S, 	& 	x < -S\\
		x, 		& 	\mbox{otherwise}
	\end{cases}
\end{align}
and
the dead-zone-type nonlinearity
\begin{align}
f(x)&=
	\begin{cases}
		x-D, 		& 	x>D\\
		x+D, 	& 	x < -D\\
		0. 		& 	\mbox{otherwise}
	\end{cases}
\end{align}
Here, $S$ and $D$ are the saturation value and dead-zone width, respectively,
and they are nonnegative real numbers.
Figs. \ref{fig:2nonlinearities}(a) and (b) illustrate 
the saturation-type and 
dead-zone-type nonlinearities, respectively.
The dead-zone-type nonlinearity is also an important 
nonlinear function called the soft thresholding function. 
It is also utilized in reconstruction algorithms such as 
ISTA\cite{Chambolle1998} and FISTA\cite{Beck2009} 
for compressed sensing\cite{Claerbout1973,Candes2006,Donoho2006}.

%

The error signal, denoted as $e(n)$, is formulated 
by incorporating the independent background 
noise component $\xi(n)$ into the difference 
between $f(x(n))$ and $y(n)$. 
That is, 
\begin{align}
e(n)&=f(x(n))-y(n)+\xi(n).	\label{eqn:error}
\end{align}
In this analysis, $\xi(n)$ is characterized solely by 
a mean of zero and a variance denoted as $\sigma_\xi^2$. 
It is important to emphasize that these parameters 
do not presuppose any specific probability distribution for 
the background noise, 
including but not limited to the Gaussian distribution.

The LMS algorithm\cite{Widrow1960}
is employed to update the adaptive filter.
That is,
\begin{align}
\bm{w}(n+1)	&=	\bm{w}(n)+\mu e(n)\bm{u}(n),		\label{eqn:LMS}
\end{align}
where $\mu$ represents the step size, which is a positive real number.

\section{Analysis} \label{sec:analysis}
In this section, we conduct a theoretical analysis 
using the statistical-mechanical method to compare 
the behaviors of the adaptive system in two separate scenarios: 
one with saturation-type nonlinearity and 
the other with dead-zone-type nonlinearity.
From (\ref{eqn:error}), the MSE is expressed as
\begin{align}
\left\langle e^2\right\rangle 
&= \left\langle \left( f(x)-y+\xi\right)^2\right\rangle\\
&= \left\langle f(x)^2\right\rangle+\left\langle y^2\right\rangle
-2\left\langle f(x)y\right\rangle+\sigma_\xi^2.	\label{eqn:MSE1}
\end{align}
In this section, we omit the time step $n$
unless otherwise stated
to avoid 
cumbersome notation.
We assume 
$N \to \infty$\footnote{This is called 
{\it the thermodynamic limit} in statistical mechanics.} 
while keeping 
\begin{align}
\rho^2 &\triangleq N\sigma^2 \label{eqn:rho}
\end{align}
constant, in accordance with the statistical-mechanical
method\cite{Nishimori2001}.
The normalized LMS (NLMS) algorithm\cite{Haykin2002,Sayed2003}  
is a significant variant of the LMS algorithm. 
The update rule of the NLMS algorithm is detailed as follows:
\begin{align}
\bm{w}(n+1)	&=	\bm{w}(n)+\frac{\tilde{\mu}}{\|\bm{u}(n)\|_2^2} e(n)\bm{u}(n),		
\label{eqn:NLMS}
\end{align}
where $\tilde{\mu}$ is the step size.
Given that $\|\bm{u}(n)\|^2_2 = N\sigma^2 = \rho^2$, 
the present analysis 
is equivalent to the analysis of 
the NLMS algorithm where $\tilde{\mu}=\rho^2\mu$ serves as the step size
for a stationary input signal $u(n)$.
%
Then, according to the central limit theorem, 
both $x$ and $y$ are stochastic variables 
that follow the Gaussian distribution.
Their means are zero, and  their variance--covariance matrix is
\begin{align}
	\bm{\Sigma}&\triangleq 
	\rho^2
	\begin{pmatrix}
		\sigma_g^2 	& r		\\
		r 				& Q
	\end{pmatrix}
	\label{eqn:cov}
\end{align}
\cite{Miyoshi2022}.
Here, $r$ and $Q$ are macroscopic variables that are respectively defined as
\begin{align}
r &\triangleq 	\frac{1}{N}\bm{g}^\top\bm{w},	\label{eqn:rdef} \\
Q &\triangleq \frac{1}{N}\bm{w}^\top\bm{w}. \label{eqn:Qdef}
\end{align}
The derivation of the means and variance--covariance matrix
is given in detail in Appendix \ref{sec:app_mean_cov}.

We obtain three sample means in (\ref{eqn:MSE1}) as follows:
\begin{itemize}
\item Saturation type
\begin{align}
\lla f(x)^2\rra 
&=S^2-S\sqrt{\frac{2\rho^2\sigma_g^2}{\pi}}\exp\lp -\frac{S^2}{2\rho^2\sigma_g^2}\rp  \nonumber \\
&+\lp \rho^2\sigma_g^2-S^2\rp \mbox{erf}\lp\frac{S}{\sqrt{2\rho^2 \sigma_g^2}}\rp, 
\label{eqn:Sfx2}	\\
%
\lla y^2\rra
&=\rho^2 Q,	
\label{eqn:Sy2} \\
\lla f(x)y\rra
&=\rho^2 r\ \mbox{erf}\lp\frac{S}{\sqrt{2\rho^2 \sigma_g^2}}\rp,	
\label{eqn:Sfxy}
\end{align}
\item Dead-zone type
\begin{align}
\lla f(x)^2\rra 
&=\lp D^2+\rho^2 \sigma_g^2\rp 
\lp 1-\mbox{erf}\lp\frac{D}{\sqrt{2\rho^2 \sigma_g^2}}\rp \rp \nonumber \\
&-D\sqrt{\frac{2\rho^2\sigma_g^2}{\pi}}\exp\lp -\frac{D^2}{2\rho^2\sigma_g^2}\rp, 
\label{eqn:Dfx2}	\\
\lla y^2\rra
&=\rho^2 Q,	
\label{eqn:Dy2} \\
\lla f(x)y\rra
&=\rho^2 r\ \lp 1-\mbox{erf}\lp\frac{D}{\sqrt{2\rho^2 \sigma_g^2}}\rp \rp,	
\label{eqn:Dfxy}
\end{align}
\end{itemize}
where $\mbox{erf}(\cdot)$ is an error function defined as
\begin{align}
\mbox{erf}(x)
&\triangleq \frac{2}{\sqrt{\pi}}\int_0^x \exp\left(-\tau^2\right)\mathrm{d}\tau.
\end{align}
%
Equation (\ref{eqn:Sy2}) is readily obtained from (\ref{eqn:cov}).
Equations (\ref{eqn:Sfx2}) and (\ref{eqn:Sfxy}) 
are derived in detail in Appendices \ref{sec:appSfx2} and \ref{sec:appSfxy},
respectively.
The details of the calculation of sample averages that appear 
thereafter are omitted, but they can be obtained using almost identical calculations.

%
Substituting (\ref{eqn:Sfx2})--(\ref{eqn:Sfxy}) 
and
(\ref{eqn:Dfx2})--(\ref{eqn:Dfxy})
into (\ref{eqn:MSE1}),
we obtain the MSE for the saturation and dead-zone types,
respectively.
In both scenarios, 
the MSE depends on
the macroscopic variables $r$ and $Q$.
Therefore, 
we formulate differential equations to characterize
the dynamical behaviors of these variables in the following.
%
Multiplying 
both sides of (\ref{eqn:LMS}) on the left by 
$\bm{g}^\top$
and using (\ref{eqn:rdef}), we obtain
\begin{align}
Nr(n+1) &= Nr(n) + \mu e(n)x(n). 
\label{eqn:Nr0}
\end{align}
%
We introduce 
time $t$ defined by
\begin{align}
t &\triangleq \frac{n}{N}
\end{align}
and use it to
represent the adaptive process.
Then, $t$ transitions to a continuous variable since the limit
$N\rightarrow \infty$ is considered.
This approach is consistent with the 
statistical-mechanical methods for online learning\cite{Engel2001}.

If the adaptive filter is updated 
$N\mathrm{d}t$ times in an infinitely small time $\mathrm{d}t$, 
we can obtain $N\mathrm{d}t$ equations as
\begin{align}
Nr(n+1) &= Nr(n)    + \mu e(n)x(n),       \label{eqn:Nr_n+1} \\  
Nr(n+2) &= Nr(n+1) + \mu e(n+1)x(n+1), \label{eqn:Nr_n+2} \\  
\vdots \ \ \ \ \  &\hspace{1.5mm}\vdots \ \ \ \ \ \ \ \vdots \nonumber \\
Nr(n+N\mathrm{d}t) 
    &= Nr(n+N\mathrm{d}t-1) \nonumber \\
    &\hspace{5mm} + \mu e(n+N\mathrm{d}t-1)x(n+N\mathrm{d}t-1). \label{eqn:Nr_n+Ndt}
\end{align}
Summing all these equations,
we obtain
\begin{align}
Nr(n+N\mathrm{d}t) &= Nr(n)+\mu \sum_{n'=n}^{n+N\mathrm{d}t-1}e(n')x(n').
\label{eqn:N(r+Ndt)}
\end{align}
Therefore, we obtain
\begin{align}
N(r+\mathrm{d}r) &= Nr + N\mathrm{d}t \mu \lla ex\rra. \label{eqn:Nr+dr}
\end{align}
%
Here, drawing from the law of large numbers, 
we have represented the effect of probabilistic variables through 
their means, considering that updates occur $N\mathrm{d}t$ times
-- numerous occasions that result in the increase in $r$ by $\mathrm{d}r$.
This property 
is called {\it self-averaging} in statistical mechanics\cite{Nishimori2001}.
From (\ref{eqn:error}) and (\ref{eqn:Nr+dr}), 
we derive a differential equation that systematically outlines 
the deterministic dynamics of $r$ as follows:
%
\begin{align}
\frac{\mathrm{d}r}{\mathrm{d}t}
&=\mu \Bigl( \lla f(x)x \rra - \lla xy\rra \Bigr). 
\label{eqn:drdt} 
\end{align}
Next, by squaring both sides of (\ref{eqn:LMS}) 
and applying the method used to derive the differential equation for 
$r$, we successfully derive a differential equation for $Q$. 
The derived equation is expressed as
\begin{align}
\frac{\mathrm{d}Q}{\mathrm{d}t}
&=\mu^2 \rho^2 \Bigl( \lla f(x)^2\rra - 2\lla f(x)y\rra
+\lla y^2\rra+\sigma_\xi^2\Bigr) \nonumber \\
&+2\mu \Bigl( \lla f(x)y\rra - \lla y^2\rra \Bigr).
\label{eqn:dQdt}
\end{align}
In this analysis, we leverage the fact that as $N\rightarrow \infty$, 
the term $\mathbf{u}^\top \mathbf{u}=\|\mathbf{u}\|_2^2=\sum_{i=1}^N u(n-i+1)^2$
transitions from a random variable to a constant value $\rho^2=N\sigma^2$.
This transition is supported by the law of large numbers, which allows
$\mathbf{u}^\top \mathbf{u}$ 
to be consistently treated as a constant outside the expectation operation
$\langle \cdot \rangle$.
This approach highlights a significant advantage of 
the statistical-mechanical method in scenarios where
$N\rightarrow \infty$, enabling simplifications 
that are not apparent under finite conditions.


Equations (\ref{eqn:drdt}) and (\ref{eqn:dQdt}) include five sample means.
However, because three of the five means are already given in 
(\ref{eqn:Sfx2})--(\ref{eqn:Sfxy})
or
(\ref{eqn:Dfx2})--(\ref{eqn:Dfxy}),
we similarly obtain the two remaining means as follows:
\begin{itemize}
%
%
\item{Saturation type}
\begin{align}
\lla xy\rra 
&=\rho^2 r, \label{eqn:Sxy} \\
\lla f(x)x\rra
&=\rho^2 \sigma_g^2\ \mbox{erf}\lp\frac{S}{\sqrt{2\rho^2 \sigma_g^2}}\rp,	
\label{eqn:Sfxx}
\end{align}
%
%
\item{Dead-zone type}
\begin{align}
\lla xy\rra 
&=\rho^2 r, \label{eqn:Dxy} \\
\lla f(x)x\rra
&=\rho^2 \sigma_g^2\ \lp 1-\mbox{erf}\lp\frac{D}{\sqrt{2\rho^2 \sigma_g^2}}\rp \rp.	
\label{eqn:Dfxx}
\end{align}
\end{itemize}

Substituting 
(\ref{eqn:Sfx2})--(\ref{eqn:Sfxy}), (\ref{eqn:Sxy}), and (\ref{eqn:Sfxx})
or
(\ref{eqn:Dfx2})--(\ref{eqn:Dfxy}), (\ref{eqn:Dxy}), and (\ref{eqn:Dfxx})
into (\ref{eqn:drdt}) and (\ref{eqn:dQdt}),
we obtain the concrete formulas of the simultaneous differential equations 
for the saturation or dead-zone type.
%
%
By analytically solving the derived simultaneous differential equations, 
we can explicitly obtain the macroscopic variables as follows:
\begin{itemize}
\item{Saturation type}
\begin{align}
\hspace{-5mm}
r(t)&=
\sigma_g^2 \mbox{ erf}\lp \frac{S}{\sqrt{2 \rho^2 \sigma_g^2}} \rp 
\Big( 1-\exp\lp -\mu \rho^2 t\rp \Big),
\label{eqn:Sr}\\
Q(t)&=
-2\sigma_g^2 \mbox{ erf}\lp \frac{S}{\sqrt{2 \rho^2 \sigma_g^2}} \rp^2
\exp\lp -\mu \rho^2 t\rp \nonumber \\
&+ A_S \exp\lp -\mu \rho^2 \lp 2-\mu \rho^2\rp t\rp
\nonumber \\
&+ 2\sigma_g^2 
\mbox{ erf}\lp \frac{S}{\sqrt{2 \rho^2 \sigma_g^2}} \rp^2 
-A_S,
\label{eqn:SQ} 
\end{align}
\begin{align}
\hspace{-5mm}
A_S&=
\frac{1}{2-\mu \rho^2}
\Biggl[ 2\sigma_g^2 \mbox{ erf}\lp \frac{S}{\sqrt{2 \rho^2 \sigma_g^2}} \rp^2
\nonumber \\
&- \mu \lp S^2 \lp 1-\mbox{ erf}\lp \frac{S}{\sqrt{2 \rho^2 \sigma_g^2}} \rp \rp \right. \nonumber \\
&+ \rho^2 \sigma_g^2 \mbox{ erf}\lp \frac{S}{\sqrt{2 \rho^2 \sigma_g^2}} \rp
\nonumber \\
&- \left.  S\sqrt{\frac{2 \rho^2 \sigma_g^2}{\pi}}\exp \lp -\frac{S^2}{2 \rho^2 \sigma_g^2}\rp
+\sigma_\xi^2\rp \Biggr],
\label{eqn:AS}
\end{align}
\item{Dead-zone type}
\begin{align}
\hspace{-5mm}
r(t)&=
\sigma_g^2 \lp 1-\mbox{ erf}\lp \frac{D}{\sqrt{2 \rho^2 \sigma_g^2}} \rp \rp
 \nonumber \\
&\times \Bigg( 1-\exp \Big( -\mu \rho^2 t\Big) \Bigg),
\label{eqn:Dr} \\
Q(t)&=
-2\sigma_g^2 \lp 1- \mbox{ erf}\lp \frac{D}{\sqrt{2 \rho^2 \sigma_g^2}} \rp \rp^2
\exp \Big(-\mu \rho^2 t\Big) \nonumber \\
&+ A_D \exp \Big( -\mu \rho^2 (2-\mu \rho^2) t\Big) \nonumber \\
&+ 2\sigma_g^2 \lp 1-\mbox{ erf}\lp \frac{D}{\sqrt{2 \rho^2 \sigma_g^2}} \rp\rp^2 -A_D,
\label{eqn:DQ}\\
A_D&=
\frac{1}{2-\mu \rho^2}
\Biggl[ 2\sigma_g^2 \lp 1-\mbox{ erf}\lp \frac{D}{\sqrt{2 \rho^2 \sigma_g^2}} \rp \rp^2
\nonumber \\
&- \mu \lp D^2 \lp 1-\mbox{ erf}\lp \frac{D}{\sqrt{2 \rho^2 \sigma_g^2}} \rp \rp \right. \nonumber \\
&+ \rho^2 \sigma_g^2 \lp 1-\mbox{ erf}\lp \frac{D}{\sqrt{2 \rho^2 \sigma_g^2}} \rp \rp
\nonumber \\
&- \left.  D\sqrt{\frac{2 \rho^2 \sigma_g^2}{\pi}}\exp \lp -\frac{D^2}{2 \rho^2 \sigma_g^2}\rp
+\sigma_\xi^2\rp \Biggr].
\label{eqn:AD}
\end{align}
\end{itemize}

Note that the simultaneous differential equations 
obtained by 
substituting 
(\ref{eqn:Sfx2})--(\ref{eqn:Sfxy}), (\ref{eqn:Sxy}), and (\ref{eqn:Sfxx})
or
(\ref{eqn:Dfx2})--(\ref{eqn:Dfxy}), (\ref{eqn:Dxy}), and (\ref{eqn:Dfxx})
into (\ref{eqn:drdt}) and (\ref{eqn:dQdt})
can be solved analytically, 
since their right-hand sides are linear expressions for $r$ and $Q$. 
This is in contrast to our group's analyses for active noise control
\cite{Miyoshi2011EL,Miyoshi2018,Miyoshi2022}, 
where the differential equation could not be solved analytically 
and had to be solved numerically.

Upon closer examination of the terms involving time $t$ in 
(\ref{eqn:Sr}), (\ref{eqn:SQ}), (\ref{eqn:Dr}), and (\ref{eqn:DQ}),
it becomes evident that $r$ and $Q$ exhibit 
two distinct time constants, that is, $(\mu \rho^2)^{-1}$ and $(\mu \rho^2(2-\mu \rho^2))^{-1}$.
Substituting the obtained analytical solution 
(\ref{eqn:Sr})--(\ref{eqn:AS}) into 
(\ref{eqn:MSE1}) and (\ref{eqn:Sfx2})--(\ref{eqn:Sfxy})
yields the analytical MSE for the saturation type, 
whereas substituting 
(\ref{eqn:Dr})--(\ref{eqn:AD}) into 
(\ref{eqn:MSE1}) and (\ref{eqn:Dfx2})--(\ref{eqn:Dfxy})
yields the analytical MSE for the dead-zone type
as follows:
\begin{itemize}
\item{Saturation type}
\begin{align}
\hspace{-5mm}
\mbox{MSE}(t)&=
\rho^2 A_S \exp\lp -\mu \rho^2 (2-\mu \rho^2) t\rp \nonumber \\
&+
\frac{2\rho^2\sigma_g^2}{2-\mu \rho^2}
\Bigg[
\lp 1-\mbox{erf}\lp\frac{S}{\sqrt{2\rho^2 \sigma_g^2}}\rp \rp
\nonumber \\
&\hspace{4mm}
\times \lp \frac{S^2}{\rho^2 \sigma_g^2} + \mbox{erf}\lp\frac{S}{\sqrt{2\rho^2 \sigma_g^2}}\rp \rp
\nonumber \\
&\hspace{4mm}
-\sqrt{\frac{2S^2}{\pi \rho^2 \sigma_g^2}}\exp\lp -\frac{S^2}{2\rho^2\sigma_g^2}\rp \Bigg]
+
\frac{2}{2-\mu \rho^2}\sigma_\xi^2,
\label{eqn:SMSE}
\end{align}
\item{Dead-zone type}
\begin{align}
\hspace{-5mm}
\mbox{MSE}(t)&=
\rho^2 A_D \exp\lp -\mu \rho^2 (2-\mu \rho^2) t\rp \nonumber \\
&+
\frac{2\rho^2\sigma_g^2}{2-\mu \rho^2}
\Bigg[
\lp 1-\mbox{erf}\lp\frac{D}{\sqrt{2\rho^2 \sigma_g^2}}\rp \rp
\nonumber \\
&\hspace{4mm}
\times \lp \frac{D^2}{\rho^2 \sigma_g^2} + \mbox{erf}\lp\frac{D}{\sqrt{2\rho^2 \sigma_g^2}}\rp \rp
\nonumber \\
&\hspace{4mm}
-\sqrt{\frac{2D^2}{\pi \rho^2 \sigma_g^2}}\exp\lp -\frac{D^2}{2\rho^2\sigma_g^2}\rp \Bigg]
+
\frac{2}{2-\mu \rho^2}\sigma_\xi^2.
\label{eqn:DMSE}
\end{align}
\end{itemize}

The first terms in (\ref{eqn:SMSE}) and (\ref{eqn:DMSE})
are dependent on time $t$.
By closely examining the terms,
we apparently see that MSEs contain 
only one time constant, that is, $(\mu \rho^2(2-\mu \rho^2))^{-1}$.
The difference between $A_S$ and $A_D$ 
is the cause of the difference in learning curves
shown in Sect. \ref{sec:results_and_discussion}.
From (\ref{eqn:SMSE}) and (\ref{eqn:DMSE}), 
it is easy to see that the condition for MSE convergence is
\begin{align}
0 &< \mu <\frac{2}{\rho^2}. \label{eqn:convergence}
\end{align}
If this condition is satisfied, the first terms  
in (\ref{eqn:SMSE}) and (\ref{eqn:DMSE})
converge to zero.


From (\ref{eqn:rdef}) and (\ref{eqn:Qdef}),
we can also obtain 
the MSD, or misalignment,
as a function of the macroscopic variables $r$ and $Q$ as follows:
\begin{align}
\mbox{MSD}
&= \|\bm{g}-\bm{w}\|_2^2 \\
&= \|\bm{g}\|_2^2-2\bm{g}^\top \bm{w}+\|\bm{w}\|_2^2 \\
&= N(\sigma_g^2-2r+Q).	\label{eqn:MSD}
\end{align}
Equation (\ref{eqn:MSD}) indicates that
the MSD is proportional to the tap length $N$ 
within the context of the model discussed in this research.
Therefore, the MSD is normalized by the tap length. 
This adjusted metric is referred to as the normalized MSD.
By substituting $t \rightarrow \infty$ into 
(\ref{eqn:SMSE}) and (\ref{eqn:DMSE}),
we can obtain the first terms converging to zero if (\ref{eqn:convergence}) is satisfied
and the steady-state MSE
for the saturation or dead-zone type, respectively.

Interestingly, the steady-state MSEs for the saturation and dead-zone types,
that is, the second terms in (\ref{eqn:SMSE}) and (\ref{eqn:DMSE}),
are exactly equal to each other as follows:
\begin{align}
\mbox{MSE}(\infty)
&=
\frac{2\rho^2\sigma_g^2}{2-\mu \rho^2}
\Bigg[
\lp 1-\mbox{erf}\lp\frac{Z}{\sqrt{2\rho^2 \sigma_g^2}}\rp \rp
\nonumber \\
&\hspace{4mm}
\times \lp \frac{Z^2}{\rho^2 \sigma_g^2} 
+ \mbox{erf}\lp\frac{Z}{\sqrt{2\rho^2 \sigma_g^2}}\rp \rp
\nonumber \\
&\hspace{4mm}
-\sqrt{\frac{2Z^2}{\pi \rho^2 \sigma_g^2}}
\exp\lp -\frac{Z^2}{2\rho^2\sigma_g^2}\rp \Bigg]
+
\frac{2}{2-\mu \rho^2}\sigma_\xi^2.
\label{eqn:MSEss}
\end{align}
Here, $Z$ is $S$ or $D$ for the saturation or dead-zone type,
respectively.
Equation (\ref{eqn:MSEss}) indicates 
that the background noise is enhanced by $\frac{2}{2-\mu \rho^2}$
and 
that $Z$ is included in the form $\frac{Z}{\rho \sigma_g}$.

Putting the partial differentiation of (\ref{eqn:MSEss}) with $Z$ as zero, 
we obtain the self-consistent equation for $S$ and $D$ that maximize 
the steady-state MSE as follows:
\begin{align}
\lp 1+
\sqrt{\frac{2\rho^2 \sigma_g^2}{\pi Z^2}}
\exp\lp -\frac{Z^2}{2\rho^2\sigma_g^2}\rp \rp
\mbox{erf}\lp\frac{Z}{\sqrt{2\rho^2 \sigma_g^2}}\rp
&=1.
\label{eqn:argmaxSD}
\end{align}

It can be seen that (\ref{eqn:argmaxSD})
is an equation for $\frac{Z}{\rho \sigma_g}$.
Although this equation cannot be solved analytically,
a numerical solution reveals that the condition 
that maximizes the steady-state MSE is 
\begin{align}
\frac{Z}{\rho \sigma_g} &\simeq 0.8485.
\label{eqn:MSEmaxCondition}
\end{align}
Note that $S$ and $D$ that maximize 
the steady-state MSE depend on 
neither the step size $\mu$ nor the variance $\sigma_\xi^2$ of background noise
because (\ref{eqn:argmaxSD}) does not include them.

\section{Results and Discussion} \label{sec:results_and_discussion}
\subsection{Learning curves}
We begin by examining the validity of the theory 
by comparing theoretical calculation results with simulation results, 
focusing on the dynamic behavior of the MSE, namely, the learning curves.
Figs. \ref{fig:SaturationLC} and \ref{fig:DeadzoneLC} show 
the learning curves of the saturation and dead-zone types, respectively. 
In these figures, the thick curves are the theoretical calculation results, 
whereas the thin polygonal lines are the simulation results. 
For both theoretical calculations and computer simulations, 
we set $\rho^2 = \sigma_g^2 = 1$.
%
As described in Sect. \ref{sec:model}, 
the theory in this paper does not assume a specific distribution 
for the input signal $u(n)$; however, in the computer simulations,
the input signal $u(n)$ was generated from a Gaussian distribution.
In the theoretical calculations, the results
in Figs. \ref{fig:SaturationLC} and \ref{fig:DeadzoneLC} 
correspond exactly to
the results of calculating
(\ref{eqn:SMSE}) and (\ref{eqn:DMSE}), respectively.
In the computer simulations, the adaptive filter 
W consists of 400 taps ($N=400$), and the ensemble averages over 1000 trials are plotted.

\begin{figure}[H]
\centering
\includegraphics[width=\graphsizeB\linewidth,keepaspectratio]{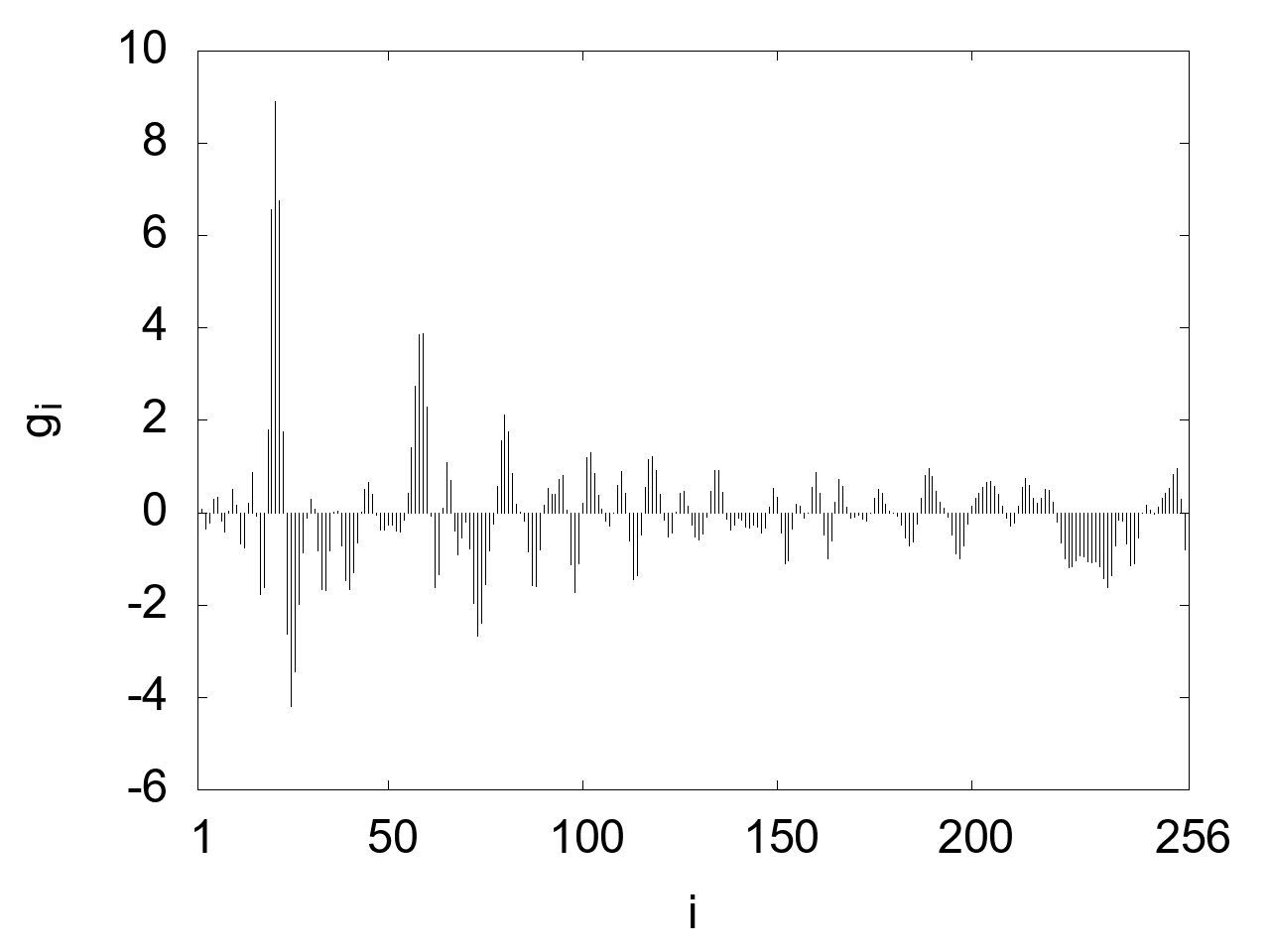}
\caption{Impulse response $\bm{g}_0$ of the unknown system G 
used in all computer simulations in this paper.}\label{fig:realunknown}
\end{figure}

The impulse response $\bm{g}_0$ of the unknown system G 
in all computer simulations in this paper was obtained from 
experimental measurements. It is depicted in Fig. \ref{fig:realunknown}.
Its dimension $M$ is 256.
Note that $\bm{g}_0$ has been normalized to meet the condition given 
in (\ref{eqn:sigmag2}).
In the simulations, all initial coefficient values $w_i(0),\ i=1,\ldots,N$ 
are set to zero. Similarly, the initial conditions 
$r(0)=Q(0)=0$ are applied in the theoretical calculations.
Figs. \ref{fig:SaturationLC} and \ref{fig:DeadzoneLC} indicate 
that the theoretical results derived in this paper 
align well with the simulation results in terms of average values.
Note, perhaps unsurprisingly, that, the learning curves for 
the saturation and dead-zone types are not the same, 
even if $S=D$.

\subsection{Steady state}
\begin{figure}[H]
    \centering
  \begin{minipage}[b]{1.00\linewidth}
    \centering
	\includegraphics[width=\graphsize\linewidth,keepaspectratio]{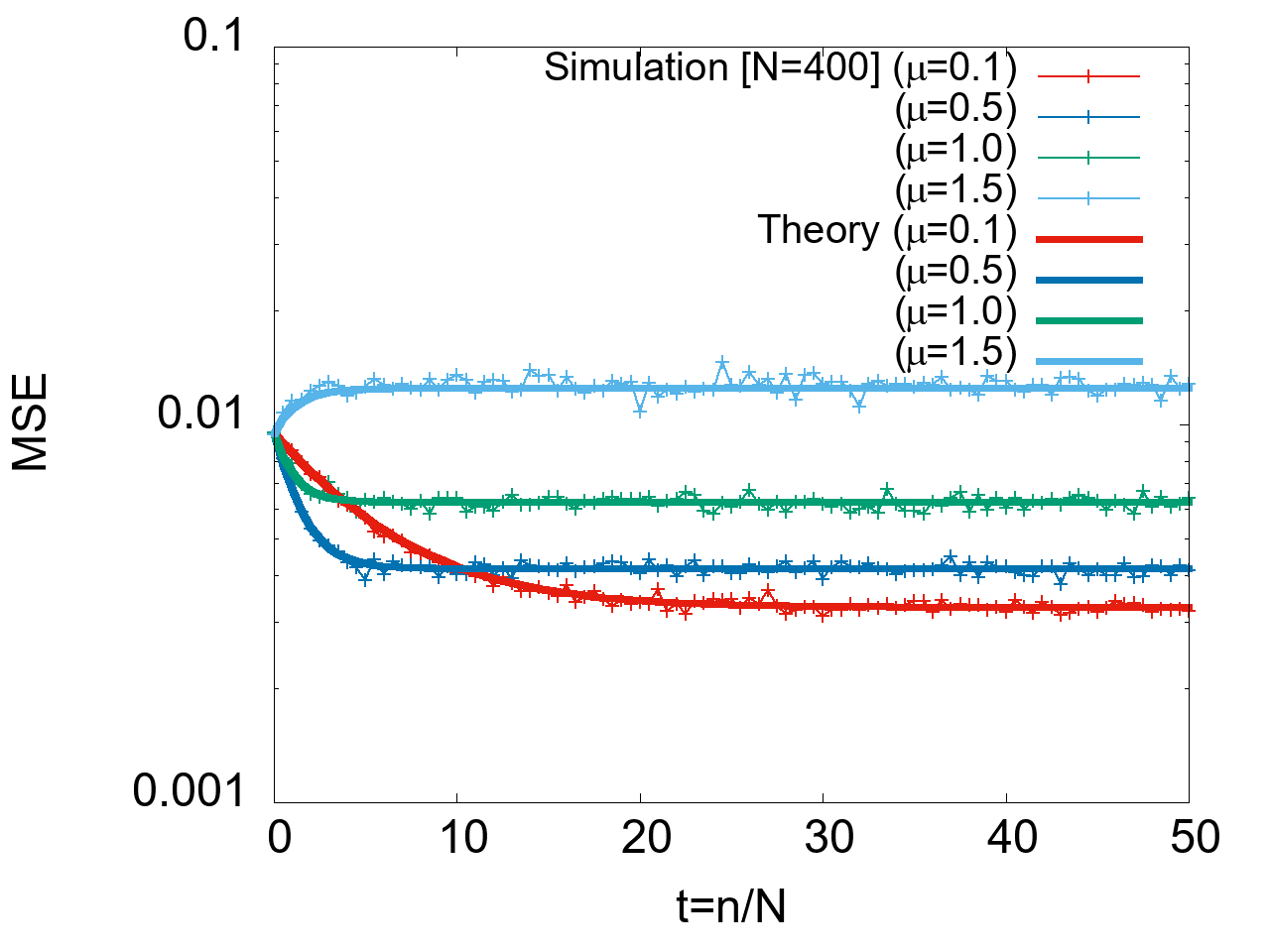}
    \subcaption{$S=0.1$}\label{fig:S01}
   \centering
    \includegraphics[width=\graphsize\linewidth,keepaspectratio]{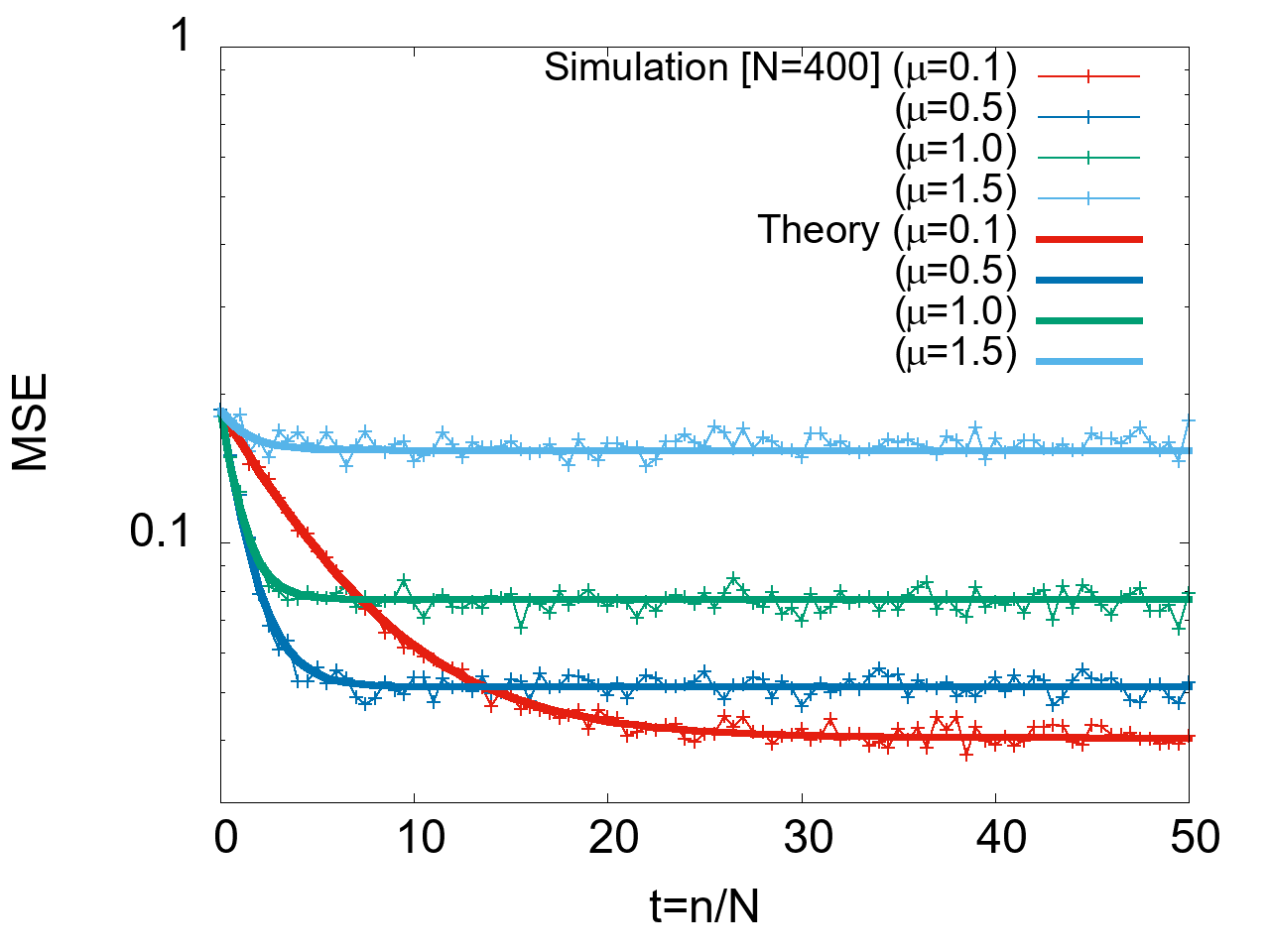}
    \subcaption{$S=0.5$}\label{fig:S05}
    \centering
	\includegraphics[width=\graphsize\linewidth,keepaspectratio]{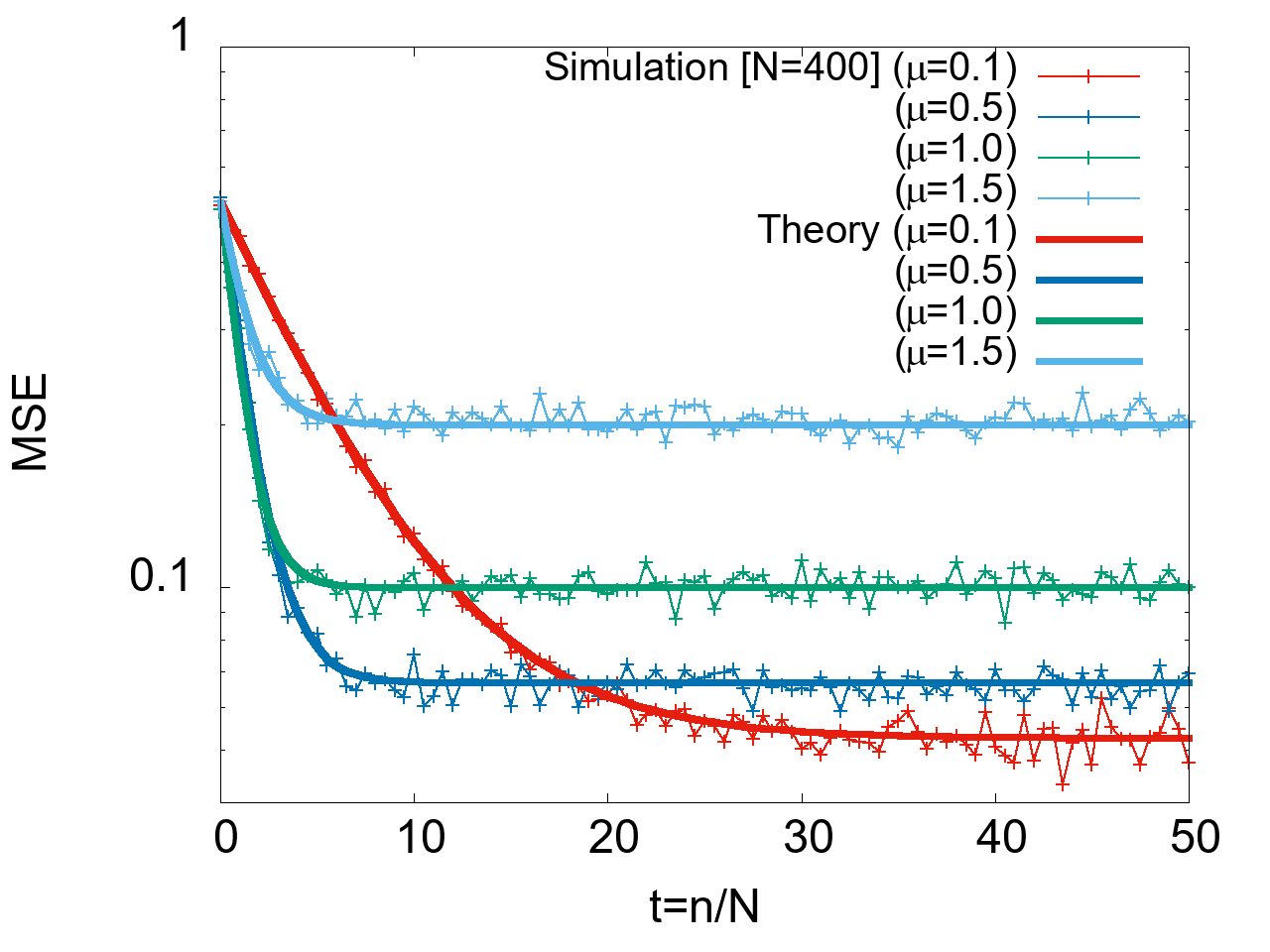}
    \subcaption{$S=1.0$}\label{fig:S10}
    \centering
	\includegraphics[width=\graphsize\linewidth,keepaspectratio]{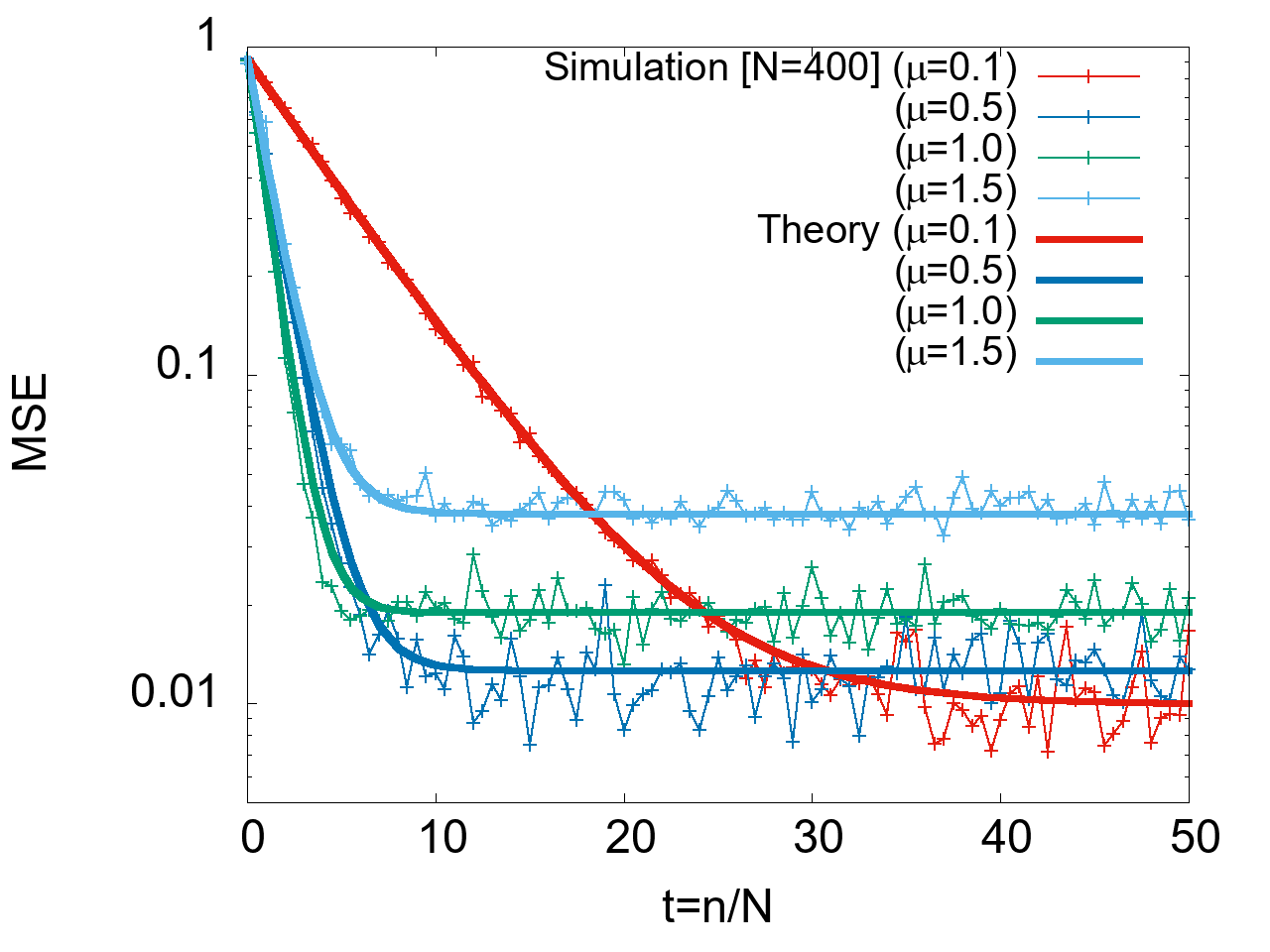}
    \subcaption{$S=2.0$}\label{fig:S20}
  \end{minipage}
%
  \caption{Learning curves (saturation type).}
  \label{fig:SaturationLC}
\end{figure}

\begin{figure}[H]
    \centering
  \begin{minipage}[b]{1.00\linewidth}
    \centering
	\includegraphics[width=\graphsize\linewidth,keepaspectratio]{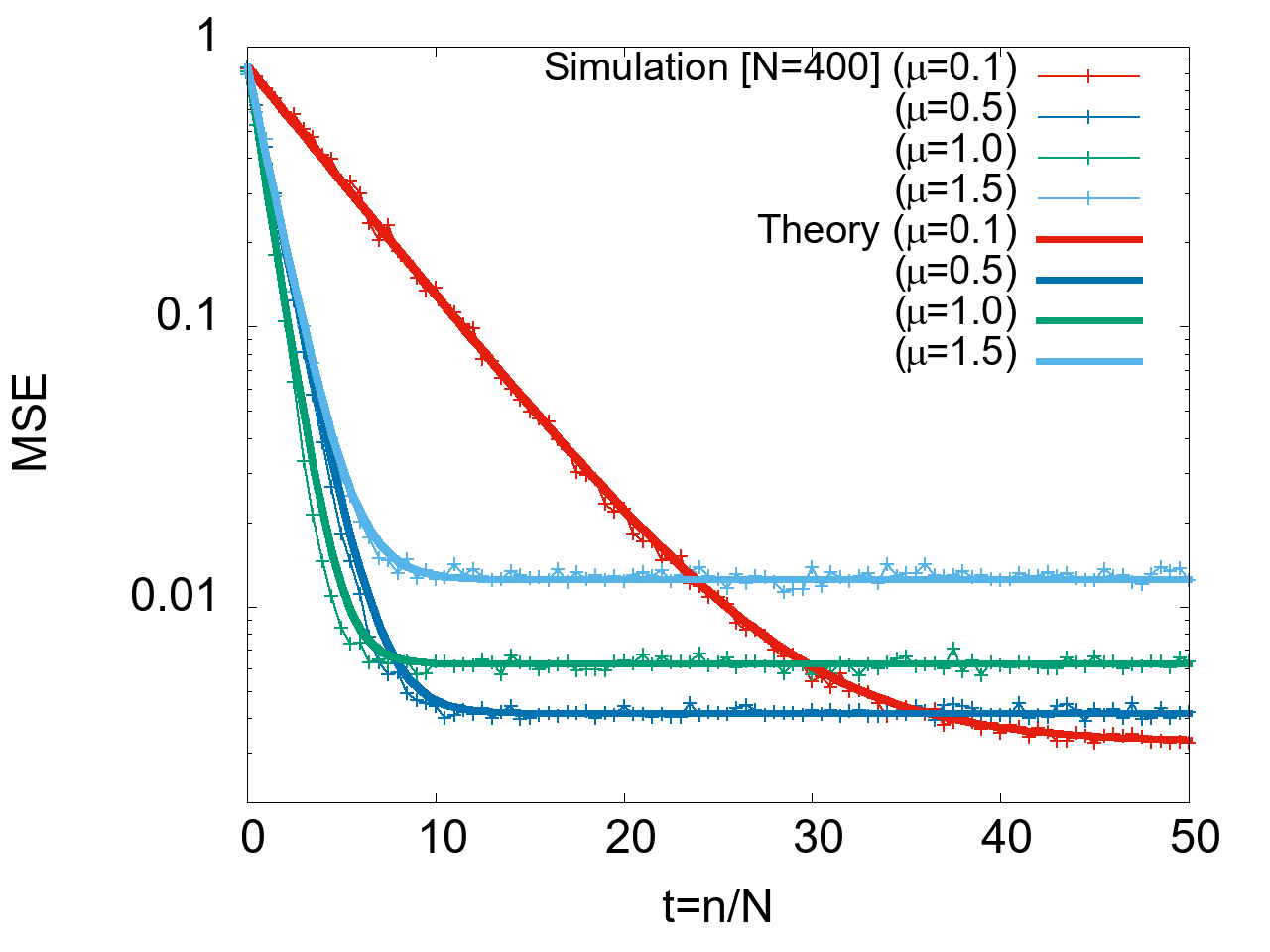}
    \subcaption{$D=0.1$}\label{fig:D01}
    \centering
    \includegraphics[width=\graphsize\linewidth,keepaspectratio]{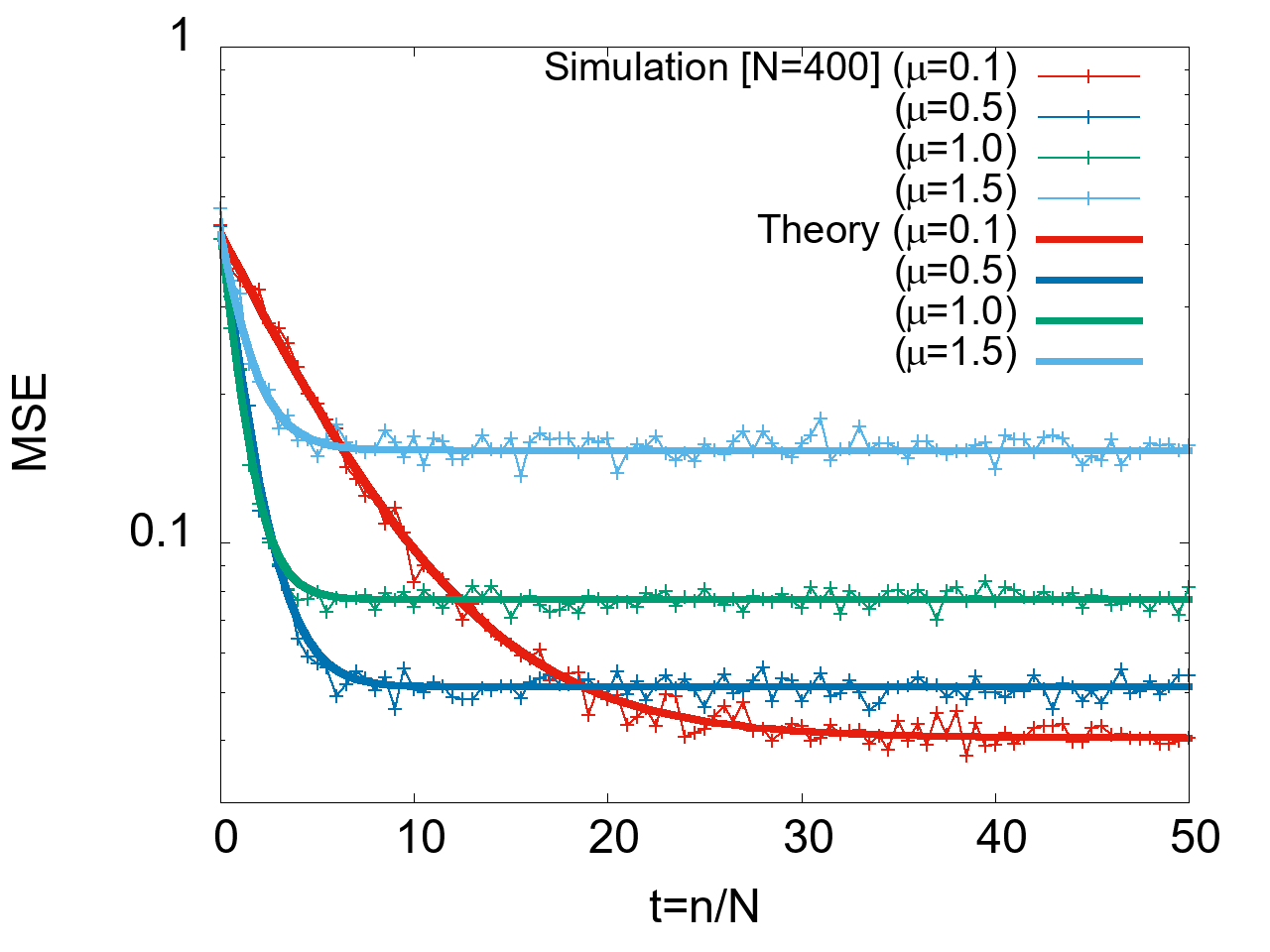}
    \subcaption{$D=0.5$}\label{fig:D05}
    \centering
	\includegraphics[width=\graphsize\linewidth,keepaspectratio]{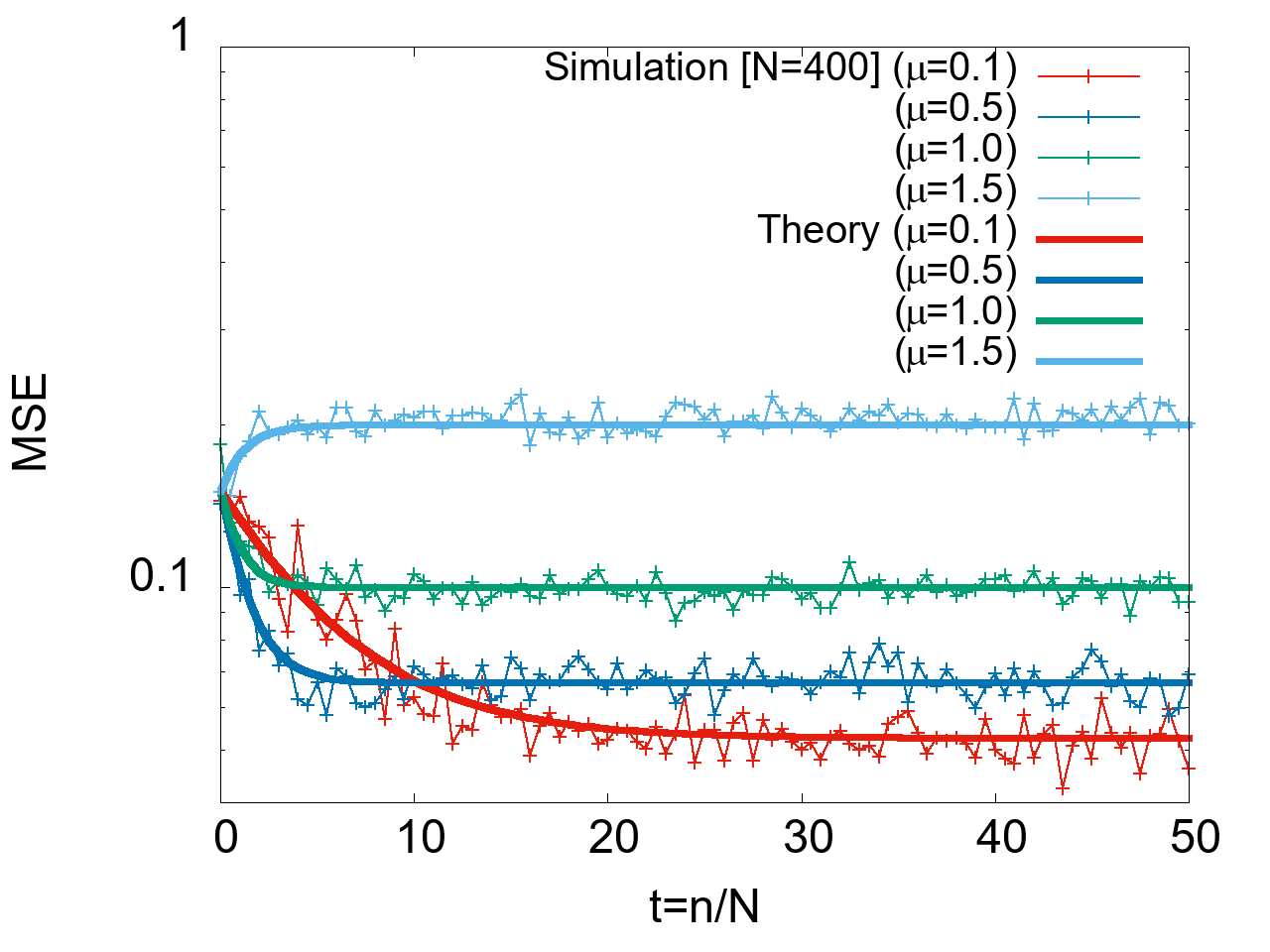}
    \subcaption{$D=1.0$}\label{fig:D10}
    \centering
	\includegraphics[width=\graphsize\linewidth,keepaspectratio]{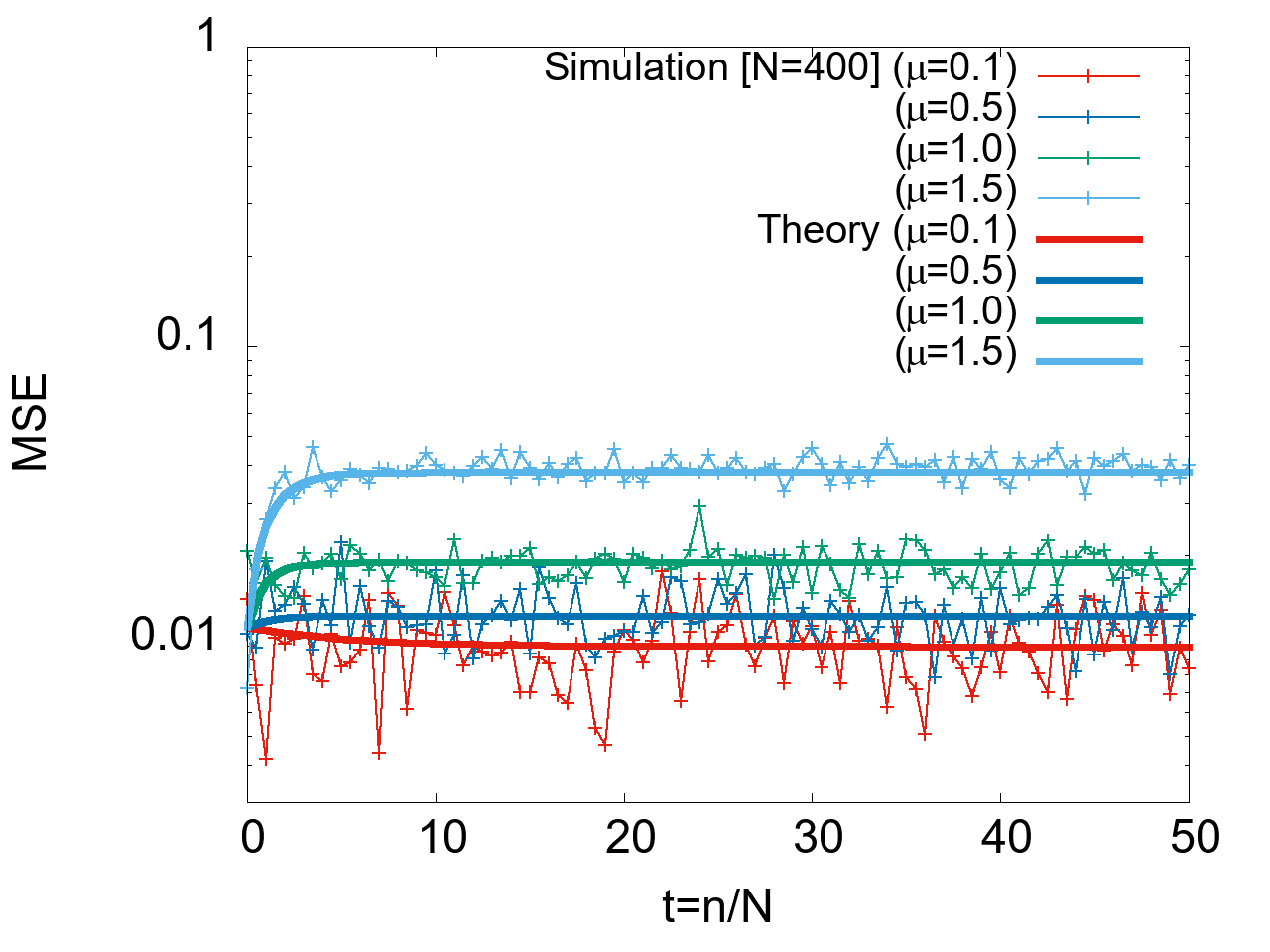}
    \subcaption{$D=2.0$}\label{fig:D20}
  \end{minipage}
%
  \caption{Learning curves (dead-zone type).}
  \label{fig:DeadzoneLC}
\end{figure}

\begin{figure}[H]
    \centering
  \begin{minipage}[b]{1.00\linewidth}
    \centering
	\includegraphics[width=\graphsize\linewidth,keepaspectratio]{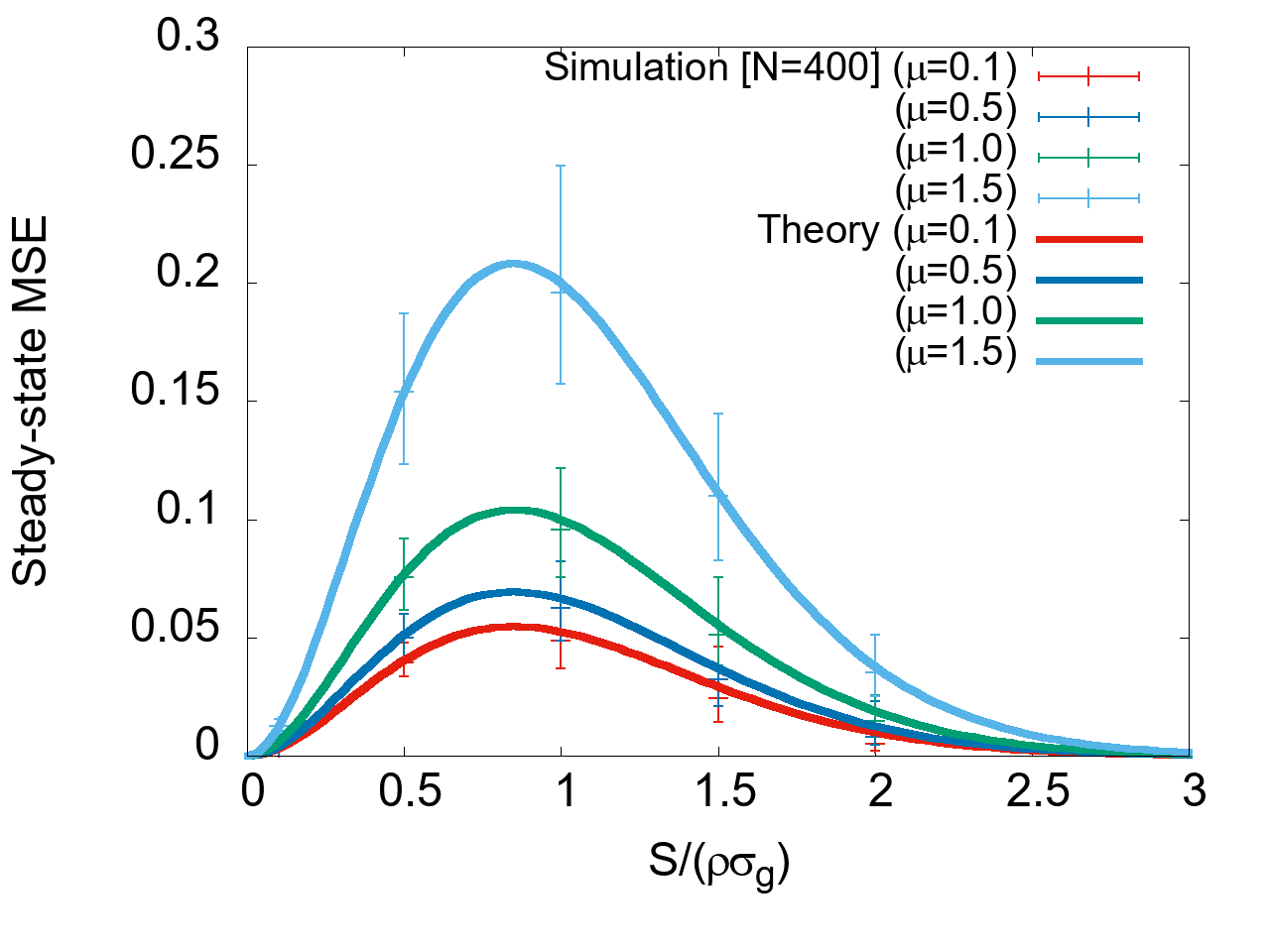}
    \subcaption{Saturation type}\label{fig:MSEssSaturation}
    \centering
    \includegraphics[width=\graphsize\linewidth,keepaspectratio]{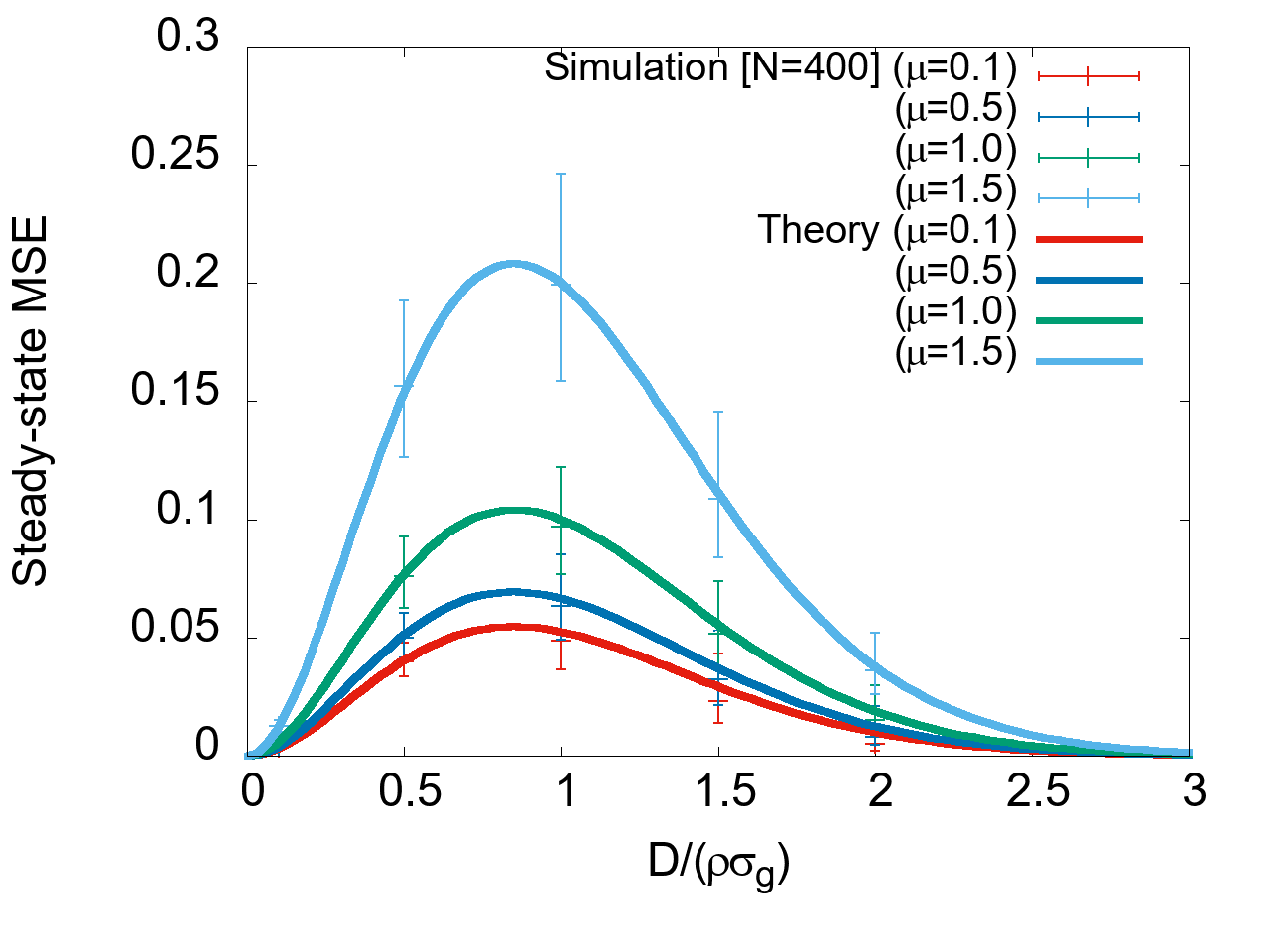}
    \subcaption{Dead-zone type}\label{fig:MSEssDeadzone}
  \end{minipage}
  \caption{Steady-state MSE.}
  \label{fig:MSEss}
\end{figure}

The learning curves in Figs. \ref{fig:SaturationLC} and \ref{fig:DeadzoneLC}
show that there appear to be steady-state values of MSE determined by 
$S$ and $\mu$ for the saturation type and by 
$D$ and $\mu$ for the dead-zone type, respectively. 
Additionally, since the MSE at $t=50$ is larger when $S$ and $D$ are 0.5 or 1 
than when they are 0.1 or 2, 
it can be inferred that the steady-state MSE is larger when $S$ or $D$ 
has intermediate values.
Thus, the steady states, like the learning curves, are 
of significant interest,
and we wish to investigate them in detail.
%
Fig. \ref{fig:MSEss} presents the 
steady-state MSE values 
obtained by the theoretical results (\ref{eqn:MSEss})
for both the saturation and dead-zone types, 
together with the corresponding simulation results at $t=200$. 
This simulation time is adequate for the MSEs to reach their steady-state values.
In the computer simulations, the adaptive filter W has 
400 taps ($N=400$).
For these simulations, error bars indicate the medians and standard deviations based on 100 trials.

Although learning curves for the saturation and dead-zone types are different, 
even if $S=D$, as shown in Figs. \ref{fig:SaturationLC} and \ref{fig:DeadzoneLC},
the steady-state MSEs for the saturation and dead-zone types 
are exactly equal to each other, as revealed in (\ref{eqn:MSEss}).
Therefore, the thick curves, that is, the theoretical calculation results, in 
Figs. \ref{fig:MSEss}(a) and (b) are, of course, identical.
It is easily seen from (\ref{eqn:MSEss}) that
if there is no background noise, that is, $\sigma_\xi^2=0$, 
the steady-state MSE is zero when $S, D=0$ or $S, D\rightarrow \infty$.
These results are also seen in Fig. \ref{fig:MSEss}.
If $S\rightarrow \infty$ or $D=0$, $f(x)$ is a linear function;
therefore, it is reasonable that the steady-state MSE is zero.
On the other hand, 
if $S=0$ or $D\rightarrow \infty$, $f(x)$ is always zero;
therefore, the output $y$ of the adaptive filter W also becomes zero
when $t \rightarrow \infty$.
As a result, the MSE becomes zero.
Here, in Fig. \ref{fig:MSEss},
the steady-state MSEs are maximum at 
$\frac{S}{\rho \sigma_g}, \frac{D}{\rho \sigma_g}\simeq 0.8485$,
as revealed in (\ref{eqn:MSEmaxCondition}).

\begin{figure}[H]
    \centering
  \begin{minipage}[b]{1.00\linewidth}
    \centering
	\includegraphics[width=\graphsize\linewidth,keepaspectratio]{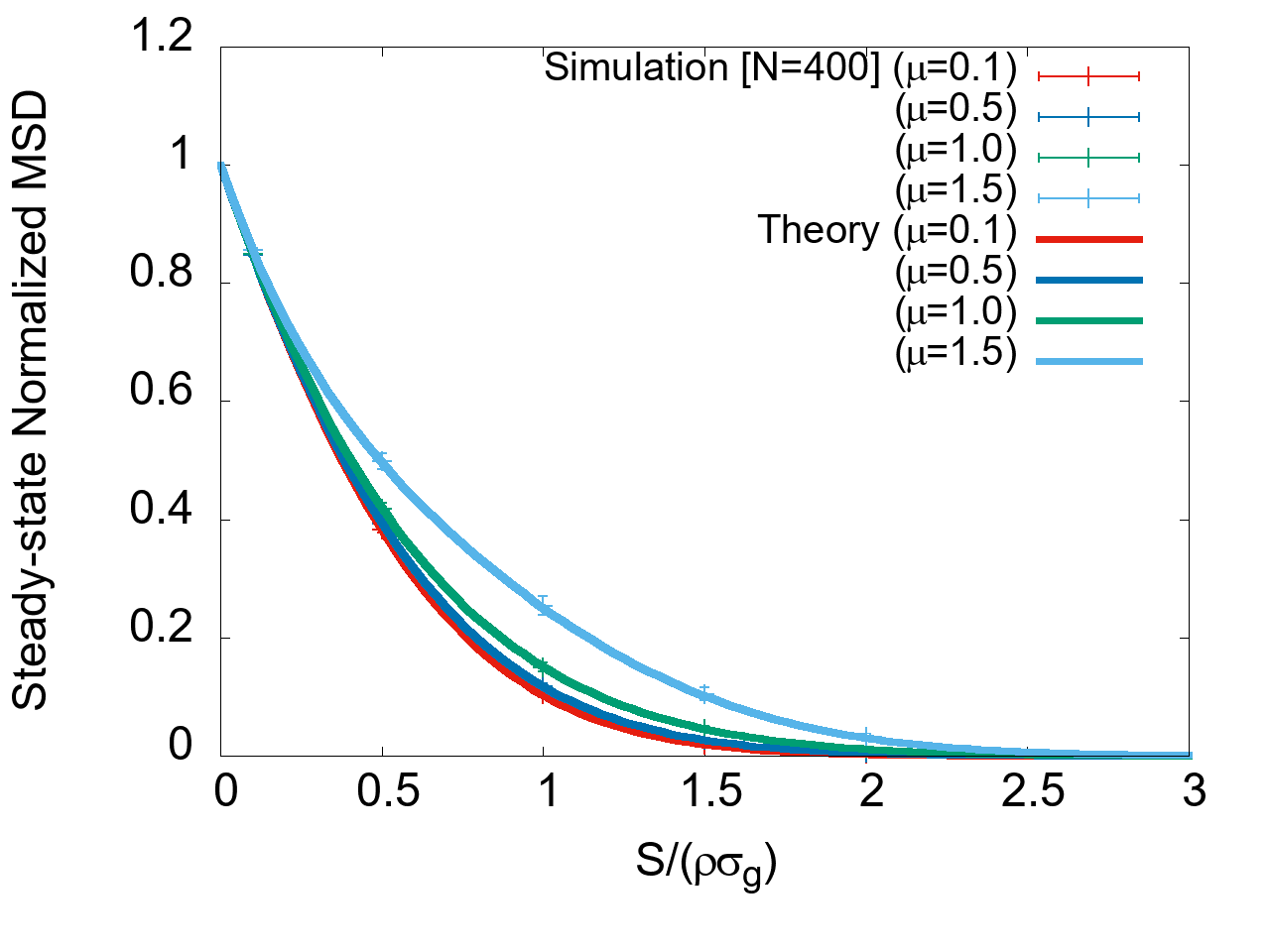}
    \subcaption{Saturation type}\label{fig:MSDssSaturation}
    \centering
	\includegraphics[width=\graphsize\linewidth,keepaspectratio]{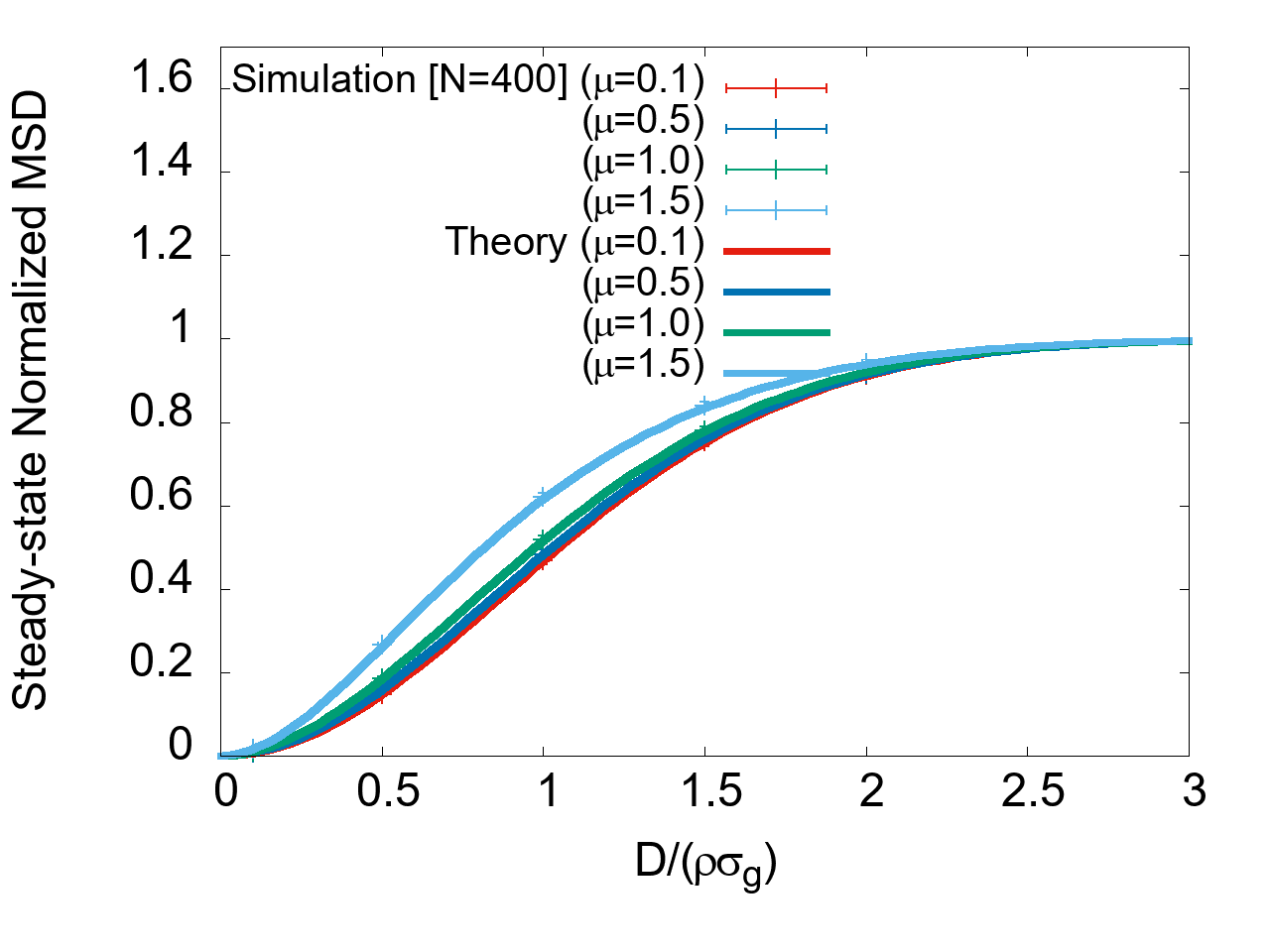}
    \subcaption{Dead-zone type}\label{fig:MSDssDeadzone}
  \end{minipage}
%
%
 \caption{Steady-state normalized MSD.}
  \label{fig:MSDss}
\end{figure}

Fig. \ref{fig:MSDss} shows the steady-state normalized MSDs 
for the saturation and dead-zone types.
Although
the steady-state MSEs for the saturation and dead-zone types 
are exactly equal to each other, as revealed in (\ref{eqn:MSEss}) 
and Fig. \ref{fig:MSEss},
the steady-state normalized MSDs are different.
That is, even if $S=D$, 
the relationship between the coefficient vector of the adaptive filter 
and that of the unknown system is not the same 
for the saturation type compared with the dead-zone type.


\section{Conclusions} \label{sec:conclusions}
In this paper, we have used the statistical-mechanical method to analyze 
the behaviors of adaptive systems with nonlinearities in 
the output of unknown systems. 
We have treated two types of nonlinearity, that is, 
the saturation-type and dead-zone-type nonlinearities. 
We have discussed the dynamical and steady-state behaviors 
of the adaptive systems.
The analysis has revealed that the steady-state MSEs of both types 
are exactly the same when the saturation value and the dead-zone width are the same. 
The self-consistent equation, which the saturation value and dead-zone width 
satisfy when the steady-state MSE is maximized, has also been obtained.
%
The theory derived in this paper allows us to predict 
how the learning curve and the steady-state MSE 
will be affected by the saturation and dead-zone properties of 
the unknown system in actual adaptive signal processing. 
In other words, the theory also provides 
important guidelines for the design of 
real-world adaptive signal processing systems 
such as acoustic echo cancellers.

The saturation-type and dead-zone-type nonlinearities
are considered highly significant, as they not only appear in 
components of practical adaptive signal processing systems, 
such as power amplifiers, loudspeakers, and microphones, 
but also in various fields of science and technology. 
Needless to say, there are many other types of nonlinearity; 
however, the analysis in this paper is also meaningful 
as a stepping stone toward addressing general nonlinearities.

A more in-depth discussion of why their steady-state MSEs 
are in exact agreement when 
the saturation value and the dead-zone width
are equal and 
determining a more general condition for the steady-state MSEs 
to be in exact agreement are some of the future issues that will be studied.
%
As described in Sect. \ref{sec:model}, 
we assumed that both the input signal and the background noise are white, but 
extending this to nonwhite cases remains a subject for future work.
Additionally, although an important feature of adaptive filters is
their ability to track time-varying unknown systems, 
their analysis presents a continuing challenge for future research.
This paper focused on the LMS algorithm, 
which is the most fundamental and important adaptive rule. 
The analysis of other adaptive algorithms, 
such as the conjugate gradient method\cite{Hestenes1952}
and 
the recursive least-squares algorithm\cite{Plackett1950,Plackett1972} 
remains a subject for future work.

\section*{Acknowledgment}
The authors wish to thank Professor Yoshinobu Kajikawa
for providing the actual data of the experimentally measured impulse responses.

\appendix
\section{Derivation of means and variance--covariance matrix of $x$ and $y$}
\label{sec:app_mean_cov}
From (\ref{eqn:sigmag2}), 
(\ref{eqn:u_mean_variance}), 
(\ref{eqn:d}), (\ref{eqn:y}), (\ref{eqn:rho}), (\ref{eqn:rdef}), 
and (\ref{eqn:Qdef}),
we obtain the means, variances, and covariance of $x$ and $y$ as follows:
\begin{align}
\lla x \rra 
&= \lla \sum_{i=1}^N g_i    u(n-i+1) \rra \\
&= \sum_{i=1}^N g_i \lla u(n-i+1) \rra =0 , \label{eqn:mean_of_d} \\
\lla y \rra 
&= \lla \sum_{i=1}^N w_i u(n-i+1) \rra \\
&= \sum_{i=1}^N w_i \lla u(n-i+1) \rra =0, \label{eqn:appy} \\
\lla x^2 \rra
&= \lla \lp \sum_{i=1}^N g_i    u(n-i+1) \rp^2 \rra \\
&= \lla \sum_{i=1}^N \sum_{j=1}^N g_i g_j   u(n-i+1) u(n-j+1) \rra \\
&= \sum_{i=1}^N g_i^2   \lla u(n-i+1)^2 \rra \\
&= \sigma^2 \sum_{i=1}^N g_i^2 
  \xrightarrow{N\rightarrow \infty} \rho^2 \sigma_g^2, 
\end{align}
\begin{align}
\lla y^2 \rra
&= \lla \lp \sum_{i=1}^N w_i    u(n-i+1) \rp^2 \rra \\
& = \lla \sum_{i=1}^N \sum_{j=1}^N w_i w_j   u(n-i+1) u(n-j+1) \rra \\
&= \sum_{i=1}^N w_i^2   \lla u(n-i+1)^2 \rra \\
&= \sigma^2 \sum_{i=1}^N w_i^2 
  \xrightarrow{N\rightarrow \infty} \rho^2 Q, \label{eqn:appy2} \\  
\lla xy \rra
&= \lla \lp \sum_{i=1}^N g_i  u(n-i+1) \rp 
          \lp \sum_{j=1}^N w_j u(n-j+1) \rp \rra  \\
&= \lla \sum_{i=1}^N \sum_{j=1}^N g_i w_j   u(n-i+1) u(n-j+1) \rra \\
&= \sum_{i=1}^N g_i w_i   \lla u(n-i+1)^2 \rra \\
&= \sigma^2 \sum_{i=1}^N g_i w_i 
  \xrightarrow{N\rightarrow \infty} \rho^2 r. \label{eqn:appdy}  
\end{align}

From (\ref{eqn:mean_of_d})--(\ref{eqn:appdy}),
the covariance matrix of $x$ and $y$ is (\ref{eqn:cov}).
Here, (\ref{eqn:appy}),  (\ref{eqn:appy2}), and  (\ref{eqn:appdy}) were derived
assuming that the correlation 
between $\bm{w}(n)$ and $\bm{u}(n)$
is small \cite{Costa2002,Tobias2000a,Tobias2000b}.  
This assumption is a standard assumption used to analyze
many adaptive algorithms\cite{Haykin2002,Sayed2003}.

\section{Derivation of (\ref{eqn:Sfx2})} \label{sec:appSfx2}
\begin{align}
\lla f(x)^2 \rra
&= \int_{-\infty}^\infty \mathrm{d}x f(x)^2 p(x) 
\\
&= \lp \int_{-\infty}^{-S} + \int_{-S}^{S}+ \int_{S}^{\infty} \rp 
   \mathrm{d}x f(x)^2 \nonumber \\
&\hspace{15mm} 
   \times \frac{1}{\sqrt{2\pi \rho^2 \sigma_g^2}}\exp \lp -\frac{x^2}{2\rho^2 \sigma_g^2}\rp 
   \\
&= 2 
   \lp
   \underbrace{\int_{S}^{\infty} \mathrm{d}x S^2 
   \frac{1}{\sqrt{2\pi \rho^2 \sigma_g^2}}
   \exp \lp -\frac{x^2}{2\rho^2 \sigma_g^2}\rp}_{\ref{sec:appSfx2}1} 
   \right.
   \nonumber \\
&\hspace{4mm}  + 
   \left.
   \underbrace{\int_0^{S} \mathrm{d}x x^2 
   \frac{1}{\sqrt{2\pi \rho^2 \sigma_g^2}}
   \exp \lp -\frac{x^2}{2\rho^2 \sigma_g^2}\rp}_{\ref{sec:appSfx2}2}
   \rp,
\end{align}

\begin{align}
\ref{sec:appSfx2}1 
&= \int_{S}^{\infty} \mathrm{d}x S^2 
   \frac{1}{\sqrt{2\pi \rho^2 \sigma_g^2}}\exp \lp -\frac{x^2}{2\rho^2 \sigma_g^2}\rp 
   \\
&=\frac{S^2}{\sqrt{\pi}} \int_{\frac{S}{\sqrt{2\rho^2 \sigma_g^2}}}^{\infty} 
  \mathrm{d}x'  \exp \lp -x'^2 \rp, \nonumber \\
&\hspace{40mm} \mbox{where } x'=\frac{x}{\sqrt{2\rho^2 \sigma_g^2}} \nonumber \\
&=\frac{S^2}{\sqrt{\pi}} 
   \lp \frac{\sqrt{\pi}}{2} -\int_0^{\frac{S}{\sqrt{2\rho^2 \sigma_g^2}}} \mathrm{d}x  \exp \lp -x^2 \rp \rp    \nonumber \\
&\hspace{30mm}  \lp \because \int_0^{\infty} \mathrm{d}x  \exp \lp -x^2 \rp = \frac{\sqrt{\pi}}{2} \rp  \nonumber \\
&=\frac{S^2}{2}\lp 1-\mbox{erf}\lp \frac{S}{\sqrt{2\rho^2 \sigma_g^2}} \rp \rp,
\end{align}

\begin{align}
\ref{sec:appSfx2}2 
&= \int_{0}^{S} \mathrm{d}xx^2 
   \frac{1}{\sqrt{2\pi \rho^2 \sigma_g^2}}\exp \lp -\frac{x^2}{2\rho^2 \sigma_g^2}\rp \\
&=\left[- x \sqrt{\frac{\rho^2 \sigma_g^2}{2\pi}}\exp \lp -\frac{x^2}{2\rho^2 \sigma_g^2} \rp\right]_0^S
\nonumber \\
&\hspace{15mm} + \sqrt{\frac{\rho^2 \sigma_g^2}{2\pi}} \int_{0}^{S} \mathrm{d}x 
   \exp \lp -\frac{x^2}{2\rho^2 \sigma_g^2}\rp, \nonumber \\
&\hspace{30mm} \mbox{where we used integration by parts } \nonumber \\
&=-S\sqrt{\frac{\rho^2 \sigma_g^2}{2\pi}}\exp \lp -\frac{S^2}{2\rho^2 \sigma_g^2} \rp
\nonumber \\
&\hspace{10mm} 
+ \frac{\rho^2 \sigma_g^2}{2} \int_{0}^{\frac{S}{\sqrt{2\rho^2 \sigma_g^2}}} \mathrm{d}x' 
   \exp \lp -x'^2\rp, 
\nonumber \\
&\hspace{45mm} 
	\mbox{where } x'=\frac{x}{\sqrt{2\rho^2 \sigma_g^2}} \nonumber \\
&=-S\sqrt{\frac{\rho^2 \sigma_g^2}{2\pi}}\exp \lp -\frac{S^2}{2\rho^2 \sigma_g^2} \rp
   + \frac{\rho^2 \sigma_g^2}{2} \mbox{erf}\lp \frac{S}{\sqrt{2\rho^2 \sigma_g^2}} \rp,  \label{eqn:J}
\end{align}

\begin{align}
\therefore 
\lla f(x)^2 \rra
&= 2 \lp \ref{sec:appSfx2}1+\ref{sec:appSfx2}2 \rp 
\\
&=S^2-S\sqrt{\frac{2 \rho^2 \sigma_g^2}{\pi}}\exp \lp -\frac{S^2}{2\rho^2 \sigma_g^2} \rp
\nonumber \\
&\hspace{10mm} 
   + \lp \rho^2 \sigma_g^2-S^2\rp \mbox{erf}\lp \frac{S}{\sqrt{2\rho^2 \sigma_g^2}} \rp.
\end{align}

\section{Derivation of (\ref{eqn:Sfxy})} \label{sec:appSfxy}
\begin{align}
\lla f(x)y \rra
&= \int_{-\infty}^\infty \int_{-\infty}^\infty 
   \mathrm{d}y \mathrm{d}x f(x)y p(x,y) \\
&= \underbrace{\int_{-\infty}^\infty \mathrm{d}y y 
   \int_{-\infty}^{-S} \mathrm{d}x (-S) p(x,y)}_{\ref{sec:appSfxy}1}
\nonumber \\
&\hspace{5mm} 
   +\underbrace{\int_{-\infty}^\infty \mathrm{d}y y \int_{-S}^S
    \mathrm{d}x xp(x,y)}_{\ref{sec:appSfxy}2}
\nonumber \\
&\hspace{10mm} 
   +\underbrace{\int_{-\infty}^\infty \mathrm{d}y y \int_S^\infty
    \mathrm{d}x S p(x,y)}_{\ref{sec:appSfxy}3},
\end{align}

\begin{align}
\ref{sec:appSfxy}2
&= \int_{-\infty}^\infty \mathrm{d}y y \int_{-S}^S
    \mathrm{d}x x 
    \frac{1}{2\pi \sqrt{\left| \rho^2 \begin{pmatrix} \sigma_g^2 & r \\ r & Q \end{pmatrix} \right|}}
\nonumber \\
&\hspace{2mm} \times
    \exp \lp -\frac{\begin{pmatrix} x & y \end{pmatrix}
    \lp \rho^2 \begin{pmatrix} \sigma_g^2 & r \\ r & Q \end{pmatrix}\rp^{-1}
    \begin{pmatrix} x \\ y \end{pmatrix}}
    {2}\rp \\
&= \int_{-\infty}^\infty \mathrm{d}y y \int_{-S}^S
    \mathrm{d}x x \frac{1}{2\pi \rho^2 \sqrt{Q \sigma_g^2 -r^2}}
\nonumber \\
&\hspace{10mm} \times
    \exp \lp -\frac{\sigma_g^2y^2-2rxy+Qx^2}
    {2\rho^2 \lp Q \sigma_g^2-r^2\rp}\rp \\
&= \int_{-\infty}^\infty \mathrm{d}y y \int_{-S}^S
    \mathrm{d}x x \frac{1}{2\pi \rho^2 \sqrt{Q \sigma_g^2 -r^2}}
\nonumber \\
&\hspace{3mm} \times
    \exp \lp -\frac{\sigma_g^2\lp y-\frac{r}{\sigma_g^2}x\rp^2
    +\lp Q-\frac{r^2}{\sigma_g^2}\rp x^2}
    {2\rho^2 \lp Q \sigma_g^2-r^2\rp}\rp \\
&= \int_{-S}^S \mathrm{d}x x 
    \exp \lp -\frac{x^2}{2\rho^2 \sigma_g^2}\rp
    \int_{-\infty}^\infty \mathrm{d}y y 
\nonumber \\
&\hspace{2mm} \times
    \frac{1}{2\pi \rho^2 \sqrt{Q \sigma_g^2 -r^2}}
    \exp \lp -\frac{\lp y-\frac{r}{\sigma_g^2}x\rp^2}
    {2\rho^2 \lp Q-\frac{r^2}{\sigma_g^2}\rp}\rp \\
&=\int_{-S}^S \mathrm{d}x x 
    \exp \lp -\frac{x^2}{2\rho^2 \sigma_g^2}\rp
    \int_{-\infty}^\infty 
    \sqrt{2\rho^2 \lp Q-\frac{r^2}{\sigma_g^2}\rp}
\nonumber \\
&\hspace{10mm} \times
    \mathrm{d}y'\lp \sqrt{2\rho^2 \lp Q-\frac{r^2}{\sigma_g^2}\rp}y'
    +\frac{r}{\sigma_g^2}x\rp 
\nonumber \\
&\hspace{10mm} \times
    \frac{1}{2\pi \rho^2 \sqrt{Q \sigma_g^2 -r^2}}
    \exp\lp -y'^2\rp, \nonumber \\
&\hspace{30mm}
    \mbox{where } y'=\frac{y-\frac{r}{\sigma_g^2}x}
	{\sqrt{2\rho^2 \lp Q-\frac{r^2}{\sigma_g^2}\rp}} \nonumber  
\end{align}
\begin{align}
&=\int_{-S}^S \mathrm{d}x x 
    \exp \lp -\frac{x^2}{2\rho^2 \sigma_g^2}\rp
    \int_{-\infty}^\infty 
    2\rho^2 \lp Q-\frac{r^2}{\sigma_g^2}\rp
\nonumber \\
&\hspace{15mm} \times
    \mathrm{d}yy\frac{1}{2\pi \rho^2 \sqrt{Q \sigma_g^2 -r^2}}
    \exp\lp -y^2\rp \nonumber \\
&\hspace{5mm}  +\int_{-S}^S \frac{r}{\sigma_g^2} \mathrm{d}x x^2 
    \exp \lp -\frac{x^2}{2\rho^2 \sigma_g^2}\rp
\nonumber \\
&\hspace{15mm} \times
    \int_{-\infty}^\infty 
    \sqrt{2\rho^2 \lp Q-\frac{r^2}{\sigma_g^2}\rp} \mathrm{d}y
\nonumber \\
&\hspace{15mm} \times
    \frac{1}{2\pi \rho^2 \sqrt{Q \sigma_g^2 -r^2}}
    \exp\lp -y^2\rp \\
&=\int_{-S}^S \frac{r}{\sigma_g^2} \mathrm{d}x x^2 
    \exp \lp -\frac{x^2}{2\rho^2 \sigma_g^2}\rp
 \nonumber \\
&\hspace{10mm} \times
   \int_{-\infty}^\infty 
    \mathrm{d}y
    \frac{1}{\pi \sqrt{2\rho^2 \sigma_g^2}}
    \exp\lp -y^2\rp \nonumber \\
&\hspace{10mm} 
   \hspace{15mm} \lp \because \int_{-S}^S \mathrm{d}x x 
    \exp \lp -\frac{x^2}{2\rho^2 \sigma_g^2}\rp=0 \rp \nonumber \\
&=\frac{2r}{\sigma_g^2\sqrt{2\pi \rho^2 \sigma_g^2}}
    \int_0^S \mathrm{d}x x^2 
    \exp \lp -\frac{x^2}{2\rho^2 \sigma_g^2}\rp
\nonumber \\
&\hspace{35mm} 
    \lp \because \int_{-\infty}^\infty \mathrm{d}y \exp\lp -y^2\rp 
    =\sqrt{\pi} \rp \nonumber \\
&=\frac{2r}{\sigma_g^2\sqrt{2\pi \rho^2 \sigma_g^2}}
    \lp \left[ -\rho^2 \sigma_g^2 x \exp \lp -\frac{x^2}{2\rho^2 \sigma_g^2}\rp \right]_0^S
\right.
\nonumber \\
&\hspace{15mm} 
\left.
    -
    \int_0^S \mathrm{d}x
    \lp -\rho^2 \sigma_g^2 \rp 
    \exp \lp -\frac{x^2}{2\rho^2 \sigma_g^2}\rp \rp, 
\nonumber \\
& \hspace{30mm} \mbox{where we used integration by parts } \nonumber \\
&=\frac{2r}{\sigma_g^2\sqrt{2\pi \rho^2 \sigma_g^2}}
    \Biggl( -\rho^2 S\sigma_g^2 \exp \lp -\frac{S^2}{2\rho^2 \sigma_g^2}\rp 
\nonumber \\
&\hspace{15mm} 
    +
    \rho^2 \sigma_g^2
    \int_0^\frac{S}{\sqrt{2\rho^2 \sigma_g^2}} 
    \sqrt{2\rho^2 \sigma_g^2}
    \mathrm{d}x'
    \exp \lp -x'^2 \rp \Biggr), 
\nonumber \\
&\hspace{45mm} 
    \mbox{where }x'=\frac{x}{\sqrt{2\rho^2 \sigma_g^2}} \nonumber \\
&=-rS\rho \sqrt{\frac{2}{\pi \sigma_g^2}}
    \exp \lp -\frac{S^2}{2\rho^2 \sigma_g^2}\rp 
    +
    \rho^2 r\  
    \mbox{erf}\lp \frac{S}{\sqrt{2\rho^2 \sigma_g^2}} \rp,
\end{align}

\begin{align}
\ref{sec:appSfxy}3
&= \int_{-\infty}^\infty \mathrm{d}y y \int_S^\infty
    \mathrm{d}x S 
    \frac{1}{2\pi \sqrt{\left| \rho^2 \begin{pmatrix} \sigma_g^2 & r \\ r & Q \end{pmatrix} \right|}}
\nonumber \\
&\hspace{2mm} \times
    \exp \lp -\frac{\begin{pmatrix} x & y \end{pmatrix}
    \lp \rho^2 \begin{pmatrix} \sigma_g^2 & r \\ r & Q \end{pmatrix}\rp^{-1}
    \begin{pmatrix} x \\ y \end{pmatrix}}
    {2}\rp \\
&= \int_{-\infty}^\infty \mathrm{d}y y \int_S^\infty
    \mathrm{d}x S \frac{1}{2\pi \rho^2 \sqrt{Q \sigma_g^2 -r^2}}
\nonumber \\
&\hspace{20mm} \times
    \exp \lp -\frac{\sigma_g^2y^2-2rxy+Qx^2}
    {2\rho^2 \lp Q \sigma_g^2-r^2\rp}\rp \\
&= \int_{-\infty}^\infty \mathrm{d}y y \int_S^\infty
    \mathrm{d}x S \frac{1}{2\pi \rho^2 \sqrt{Q \sigma_g^2 -r^2}}
\nonumber \\
&\hspace{2mm} \times
    \exp \lp -\frac{\sigma_g^2\lp y-\frac{r}{\sigma_g^2}x\rp^2
    +\lp Q-\frac{r^2}{\sigma_g^2}\rp x^2}
    {2\rho^2 \lp Q \sigma_g^2-r^2\rp}\rp \\
&= S \int_S^\infty \mathrm{d}x  
    \exp \lp -\frac{x^2}{2\rho^2 \sigma_g^2}\rp
    \int_{-\infty}^\infty \mathrm{d}y y 
\nonumber \\
&\hspace{2mm} \times
    \frac{1}{2\pi \rho^2 \sqrt{Q \sigma_g^2 -r^2}}
    \exp \lp -\frac{\lp y-\frac{r}{\sigma_g^2}x\rp^2}
    {2\rho^2 \lp Q-\frac{r^2}{\sigma_g^2}\rp}\rp \\
&=S \int_S^\infty \mathrm{d}x  
    \exp \lp -\frac{x^2}{2\rho^2 \sigma_g^2}\rp
    \int_{-\infty}^\infty 
    \sqrt{2\rho^2 \lp Q-\frac{r^2}{\sigma_g^2}\rp}
\nonumber \\
&\hspace{20mm} \times
    \mathrm{d}y'\lp \sqrt{2\rho^2 \lp Q-\frac{r^2}{\sigma_g^2}\rp}y'+\frac{r}{\sigma_g^2}x\rp
\nonumber \\
&\hspace{20mm} \times
    \frac{1}{2\pi \rho^2 \sqrt{Q \sigma_g^2 -r^2}}
    \exp\lp -y'^2\rp, \nonumber \\
&  \hspace{30mm} \mbox{where } y'=\frac{y-\frac{r}{\sigma_g^2}x}
    {\sqrt{2\rho^2 \lp Q-\frac{r^2}{\sigma_g^2}\rp}} \nonumber \\
%
&=S \int_S^\infty \mathrm{d}x 
    \exp \lp -\frac{x^2}{2\rho^2 \sigma_g^2}\rp
    \int_{-\infty}^\infty 
    2\rho^2 \lp Q-\frac{r^2}{\sigma_g^2}\rp
    \mathrm{d}yy
\nonumber \\
&\hspace{15mm} \times
    \frac{1}{2\pi \rho^2 \sqrt{Q \sigma_g^2 -r^2}}
    \exp\lp -y^2\rp \nonumber \\
&\hspace{5mm}  +S\int_S^\infty \frac{r}{\sigma_g^2} \mathrm{d}x x 
    \exp \lp -\frac{x^2}{2\rho^2 \sigma_g^2}\rp
\nonumber \\
&\hspace{15mm} \times
    \int_{-\infty}^\infty 
    \sqrt{2\rho^2 \lp Q-\frac{r^2}{\sigma_g^2}\rp} \nonumber \\
&\hspace{15mm} \times
    \mathrm{d}y
    \frac{1}{2\pi \rho^2 \sqrt{Q \sigma_g^2 -r^2}}
    \exp\lp -y^2\rp
\end{align}
\begin{align}
&=\frac{Sr}{\sigma_g^2}\int_S^\infty \mathrm{d}x x 
    \exp \lp -\frac{x^2}{2\rho^2 \sigma_g^2}\rp
    \sqrt{2\rho^2 \lp Q-\frac{r^2}{\sigma_g^2}\rp}
    \nonumber \\
&\hspace{40mm} \times
    \frac{1}{2\pi \rho^2 \sqrt{Q \sigma_g^2 -r^2}}
    \sqrt{\pi} 
    \nonumber \\ 
& \hspace{-3mm} \lp \because 
    \int_{-\infty}^\infty \mathrm{d}yy \exp\lp -y^2\rp=0,\ 
    \int_{-\infty}^\infty \mathrm{d}y \exp\lp -y^2\rp 
    =\sqrt{\pi} \rp 
    \nonumber \\
&=\frac{Sr}{\sigma_g^2}\frac{1}{\sqrt{2\pi \rho^2 \sigma_g^2}}
    \left[ \lp -\rho^2 \sigma_g^2\rp \exp \lp -\frac{x^2}{2\rho^2 \sigma_g^2}\rp \right]_S^\infty
\\
&=\frac{Sr\rho}{\sqrt{2\pi \sigma_g^2}}\exp \lp -\frac{S^2}{2\rho^2 \sigma_g^2}\rp,
\end{align}

\begin{align}
\ref{sec:appSfxy}1
&=\int_{-\infty}^\infty \mathrm{d}y y \int_{-\infty}^{-S} 
  \mathrm{d}x (-S) p(x,y)
=\ref{sec:appSfxy}3, \nonumber \\
&\hspace{2mm} \mbox{where we used the integration by substitution:}
\nonumber \\
&\hspace{45mm} 
\ x'=-x,\ y'=-y, \nonumber
\end{align}

\begin{align}
\therefore
\lla f(x)y \rra
&= \ref{sec:appSfxy}1+\ref{sec:appSfxy}2+\ref{sec:appSfxy}3
  =\rho^2 r\  
    \mbox{erf}\lp \frac{S}{\sqrt{2\rho^2 \sigma_g^2}} \rp.
\end{align}

\end{document}